\documentclass[reprint,amsmath,amssymb,aps]{revtex4-1}
\usepackage[dvipdfmx]{graphicx}
\usepackage[dvipdfmx,hidelinks]{hyperref}
\usepackage{dcolumn}
\usepackage{bm}
\usepackage{color}
\usepackage{multirow}
\usepackage{booktabs}
\usepackage{tabularx}
\usepackage{newcommand_list2} 
\begin{document}
\preprint{APS/123-QED}
\title{Resilience of antagonistic networks with regard to the effects of initial failures and degree-degree correlations}
\author{Shunsuke Watanabe}
 \email{watanabe@sp.dis.titech.ac.jp}
 \affiliation{
 Department of Computational Intelligence and Systems Science, Tokyo Institute of Technology, Yokohama 2268502, Japan
}
\author{Yoshiyuki Kabashima}%
 \email{kaba@c.titech.ac.jp}
 \affiliation{
 Department of Mathematical Intelligence and Systems Science, Tokyo Institute of Technology, Yokohama 2268502, Japan
}
\
\date{\today}
\begin{abstract}
In this study, we  investigate the resilience of duplex networked layers ($\alpha$ and $\beta$) coupled with antagonistic interlinks,  each layer of which inhibits its counterpart   at the microscopic level,
changing  the   following factors: whether the influence of the initial failures  in $\alpha$ remains (quenched (Case Q)) or not (free (Case F)); the effect of intralayer degree-degree correlations  in each layer and interlayer degree-degree correlations; and the type of the initial failures, such as random failures (RFs) or targeted attacks (TAs).
We illustrate that the percolation processes repeat in both Cases Q and F, 
although  only in Case F  are  nodes that initially failed reactivated.
To analytically evaluate the resilience of each layer, we develop a methodology based on the cavity method for deriving the size of a giant component (GC). 
Strong hysteresis, which is ignored in the standard cavity analysis, is observed in the repetition of the percolation 
processes particularly in Case F.
To handle this, we heuristically modify interlayer messages for macroscopic analysis, 
the utility of which is verified by numerical experiments. 
The percolation transition  in each layer is continuous in both Cases Q and F.
We also analyze the influences of degree-degree correlations on the robustness of layer $\alpha$, in particular for the case of TAs. 
The analysis indicates that the critical fraction of initial failures that makes the GC size in  layer $\alpha$ vanish 
depends only on its intralayer degree-degree correlations. 
Although our model is defined in a somewhat abstract manner,  it may have relevance to ecological systems that are composed of endangered species (layer $\alpha$) and invaders (layer $\beta$),  the former  of which are damaged by the latter
whereas the latter are  
exterminated  in the areas where the former are active. 
\end{abstract}
\pacs{Valid PACS appear here}
\maketitle
\section{\label{sec:level1}Introduction}
 Our real world is composed of a huge variety of systems, which function in various layers, such as technology, society, and biology, and are continuously growing in unstable environments.
It is thus of great importance to capture the essence of these complex critical systems. The {\it graph} \cite{Erdos, Bollobas} or {\it network} is one of the most powerful tools, where the constituents of the systems are regarded as nodes and the interactions between the nodes as links. 
Since it has been detected that networks representing real-world systems  exhibit small-world properties \cite{Strogatz,Watts}  and scale free (heterogeneous) properties \cite{Barabasi} in general,  various topological characteristics have been demonstrated and salient results have been reported \cite{DorogovtsevMendes,  NewmanSIAM, Boccaletti, Barthelemy}.
One of the most important  properties of a network is
its robustness, that is, its tolerance to the malfunction of some nodes and/or links, which is frequently evaluated as an aggregated property, characterized as the structural phase transition of  the emergence of a giant component (GC) \cite{Callaway}. 
Although vast  studies have been conducted in this field, research remains insufficient,  because  in most studies network patterns were projected as a single layer and  the effect exerted by the fact that real-world systems couple with one another was not realized. 
 
  For the purposes of analyzing real-world networks more essentially, the concept of {\it multilayer networks}  was developed and is considered a new paradigm of complex network science \cite{Boccaletti2, Kivela, NetofNet, Gao, Morris, Baxter, Buldyrev2011, Huang2011, Hu, Zhou2013, Shao2011, Parshani EPL, Wang, Podobnik, Huang 2013, Dong EPL2013, Cellai, Danziger2014}.  
%
The seminal work on multilayer networks is the analysis of the robustness of interdependent networks presented in \cite{Buldyrev 2010, Parshani PRL}.
A mutually connected GC (MCGC) consisting of nodes that belong to a GC in each of all the layers  collapses  even if only a portion of the nodes has initially failed in one layer,  triggering a chain of  failures (called the cascade phenomenon) that spreads over all networks.  
This type of model may in fact be the most dependable because real-world systems in general  are becoming increasingly dependent on one another \cite{Rinaldi}. 

 A different class of multiplex networks consists of those that couple with each other with antagonistic interlayer interactions, which are called antagonistic networks.   
Several papers were published with regard to robustness of antagonistic networks with neutral degree-degree correlations \cite{KZhao1, KZhao2, Kotnis}.
The theoretical framework was presented  for analyzing the robustness on duplex antagonistic networks without initial failures in \cite{KZhao1} and   later  extended to include the  failures in \cite{Boccaletti2}.
Although it was of surprise that these models exhibited the first order transition in the GC size,  
the definition of the antagonistic property of interlinks was artificial to some extent,  in particular for numerical experiments.
 In our model  the property of interlinks is  defined simply at microscopic level: nodes that belong to the GC deactivate their replica nodes, while the other nodes, which do not belong to the GC, activate  their replica nodes.  
In addition, we  explicitly define that  initial failures occur in layer $\alpha$.
 
 Although our model is defined in a somewhat abstract manner,  
 one may be able to regard it as a family of graph models for  ecological systems \cite{habitatmosaics, correlatedhabitat,habitatpercolation,conservation,spatialhabitat}.
Employing duplex networks instead of a single network, we represent  habitat patches of two categories of species (endangered species and the invasive ones) and  interactions in and between them;  
the habitats of endangered species and those of invaders are projected on  
layer $\alpha$  and layer $\beta$ respectively and the GC in each layer represents the largest and  most significant habitat of the relevant layer. Each interlayer link represents the  antagonistic interaction 
because the invaders prey on  the endangered species,  while the latter are conserved, thus  eradication program expels the former.

In this paper, we analyze the resilience of  antagonistic duplex networks  that suffer from  failures, in terms of the following three factors:
 (i) the type of the initial failures;  (ii) the remaining effect of the initial failures; and (iii) degree-degree correlations.
Two scenarios are examined featuring the two types of initial node failures in the the first layer $\alpha$:  nodes randomly  fail (RFs) or high degree nodes selectively fail,  called  targeted attacks (TAs).  The result of the initial failures propagate to the  confronting layer 
$\beta$, which causes  node failures at the second stage and  the outcome return to the layer $\alpha$. 
At the third stage, two possibilities are considered  for the remaining effect of the  initial damage. 
In one scenario, which is referred to as the {\em quenched} setting (Case Q), the effect of the initial damage remains, such that  failed nodes cannot become active again. 
 In contrast, in the second scenario, termed the {\em free} setting (Case F), all the nodes are free of the initial damage, which also implies that the nodes can be reactivated with the aid of replica nodes.

In  the above-described realistic scenario, Case F corresponds to the situation in which endangered species can recover, while in Case Q, they cannot even though invaders disappear in the area. 
In both cases  percolation processes exhibit periodic phenomena, which were  also reported in different model \cite{Kotnis}.
In addition,  we  consider the effects of two types of degree-degree correlations, those between nodes within a layer (intralayer degree-degree correlations) and those between replica nodes (interlayer degree-degree correlations). 
In general,  degrees in real-world networks are correlated \cite{Newman2002, Maslov,Newman2003}, and therefore, the influence of degree-degree correlations is considered one of the most important topics in the research on multilayer networks \cite{Boccaletti2, Zhou2012, Min}. 

As the main part of this paper, we address the development of an analytical framework based on the cavity method developed in statistical mechanics \cite{Pearl, Hartmann, Mezard, Us}, which is categorized as a mean field approach \cite{Son2012, Boccaletti2, Newman2001},  supposing a locally tree-like structure and utilizing Bethe-Peierls approximation. 
 In our framework, we first describe  the flow for computing GC size from a microscopic viewpoint, the formulation 
 of which is extended to a macroscopic viewpoint and the expected GC size is  analytically evaluated solving a set of 
 self-consistent equations numerically.
Unfortunately, the results obtained in this fashion deviate from numerical ones,  in particular in  Case F. 
The cause of the discrepancy lies in the assumption of  self-averaging property that nodes of the same degree have equivalent statistical property though at each single instance  local states of some nodes, which are  affected from the hysteresis  in 
the layer, deterministically contribute to global property of the relevant layer, namely GC size.  
The technique to efface the influence of the hysteresis is implicitly used for robustness analysis of multilayer interdependent networks \cite{Boccaletti2} and antagonistic networks \cite{KZhao1}. 
However,  this 
is not valid in  the latter case, 
unlike the former case: some inactive nodes  are regarded active, which may makes some of them belong to the GC, if it exists.
Although the fraction of these nodes is almost negligible \cite{we},  their existence may significantly influence 
the critical behavior of the system, namely  whether the manner of the percolation transition is continuous \cite{Kotnis} or discontinuous \cite{Boccaletti, KZhao1}.
In keeping with the periodic phenomena,  we heuristically describe the GC size at the microscopic level and extend it to the macroscopic one, the utility of which is confirmed  comparing  with the numerical ones. 
The percolation transition of each layer turns out to be continuous in both Cases Q and F in our model; in particular, 
that of layer $\alpha$ depends on the first stage and  the critical point is determined only by intralayer topologies in layer $\alpha$. 
On the other hand, the GC size depends on both interlayer and intralayer correlations, 
in particular the GC size in layer $\alpha$ exceeds about half of the layer size.  

The remainder of this paper is organized as follows.
In Sec. \ref{section: model}, we present the problem set-up and introduce various notations that are used in our analysis. 
In Sec. \ref{section: framework}, we  develop an analytical framework for evaluating the robustness on antagonistic networks.
In Sec. \ref{section: numerical_test}, we examine the accuracy in evaluating the GC size of our methodology. 
We find discrepancies between theory and experiment for the GC size evaluation particularly in Case F. 
For resolving this inconsistency, 
we heuristically improve the developed methodology, which is verified by numerical experiments 
in Sec. \ref{section: improvement}. 
In Sec. \ref{section: result}, we discuss the influence of interlayer and intralayer degree-degree correlations on robustness of each layer and suggest the relevance of real world ecological systems that are reported recently.
In the final section, we conclude the paper with a summary.
The periodic phenomena are reconfirmed by the heuristic and  
used notations are listed for convenience in Appendix \ref{section: reconfirming} and \ref{section: notations}, respectively.
\section{Model}
\label{section: model} 
In this section, we present a brief outline of our model of antagonistic networks consisting of layers (networks) $\alpha$ and $\beta$, where the number of nodes in each layer is $N$. They are generated separately in  some initial configuration,  where no isolated node exists in either network prior to the failure process.  

 Our model is seeded by initial damages  that destroy a portion of the nodes in layer $\alpha$, chosen uniformly at random or targeted (selected degree-dependent randomly) with probability $1-q$.  
As the result of the  first stage, a GC  may remain  in layer $\alpha$, the order of size of which is typically $O(N)$ (or $O(N^{2/3})$ at the critical point exactly) \cite{Erdos}. 
  We define that  nodes that belong to the GC in layer $\alpha$ deactivate their replica nodes in layer $\beta$, while all the other nodes that do not belong to the GC activate their replica nodes 
 which causes the failure of nodes in layer $\beta$ at the second stage, resulting in a GC in layer $\beta$ that differs from the GC in layer $\alpha$.  
 Similarly, all the nodes  that belong to the GC in layer $\beta$ deactivate their replica nodes, while the rest of nodes activate their replica nodes. 
The difference between Cases Q and F corresponds to  whether  the initial damages remain or not in layer $\alpha$ at the third stage: In Case Q, nodes are affected from  both the  initial damages and layer $\beta$, while in Case F, nodes are free from the initial damage and  only affected from layer $\beta$.

The organization of this section is as follows. In Sec. \ref{section: process},  
we  show that the failure process  oscillates in both Cases Q and F.
In Sec. \ref{section: bipartite}, we provide the bipartite graph representations of the original networks,  which are necessary for  microscopic analysis in Sec. \ref{section: micro}.
In Sec. \ref{section: topologies}, 
the topologies of each networked layer, degree distribution, and interlayer and intralayer degree-degree correlations are introduced. 
They are used for macroscopic analysis in Sec. \ref{section: macro}.   

\subsection{Percolation process}
\label{section: process}
\begin{figure}[htbp]
\centering
\includegraphics[width=8cm]{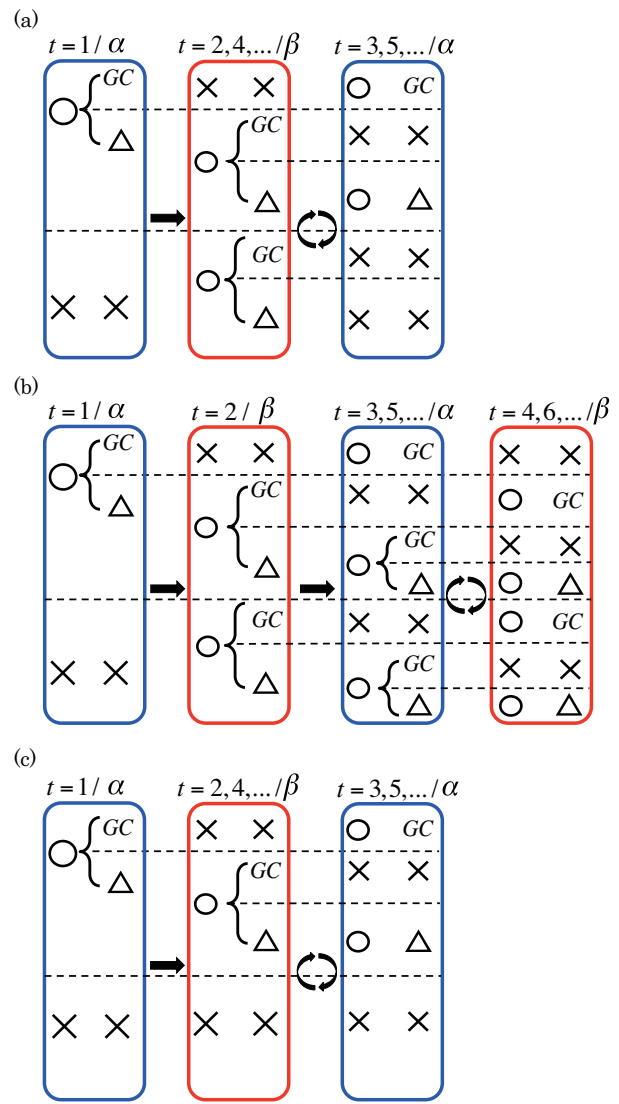}
\caption{
(Color on-line) 
Transitions of the states of nodes of the percolation process  in antagonistic networks. Each stage is expressed as a large rounded rectangle, where  red and blue represent networked layers $\alpha$ and $\beta$, respectively.
Symbols on the left hand side in a large rectangle  express the condition of the set of nodes, and  those on the right hand side represent  the percolation result  under the condition of the left hand side. 
A cross represents the set of failed nodes  at the stage, while a circle represents the set of non-failed nodes,
which are classified into two classes:
(i) nodes belonging to the GC and (ii) nodes  belonging to  one of the small components, represented by a triangular shape.  
Dotted lines separate the groups of nodes that have different percolation results.
In Case Q, only the nodes that form the GC  affect their replica nodes. In Case F, all nodes influence their replica nodes. 
(a) and (b) Possible transitions for Cases Q and F, respectively. 
(c) State transitions realized in the model examined in the case that node $\iB$ is supposed to be the replica node of node $\iA$ \cite{Boccaletti2}.}
\label{transition}
\end{figure}

In Fig. \ref{transition}, we categorize the nodes in each layer into three groups and depict them as  the stage  elapses, 
which does not depend on the initial failure type (RFs or TAs) or  any degree-degree correlations in and/or between networks. 
\begin{itemize}
\item[i)] After the $t=1$ percolation process, the nodes in layer $\alpha$ 
that constitute a GC make their replica nodes inactive at the start of stage $t=2$.
This guarantees that the nodes in layer $\alpha$ are  active at the start of stage $t=3$. 
In addition, the network topology is unchanged from stage $t=1$. 
Therefore, it is ensured that the nodes belong to the GC after the $t=3$ percolation process, and 
repeating this argument concludes that the nodes continue to constitute the GC for ever at stages $t=5, 7, \ldots$. 
Accordingly, their replica nodes continue to be inactive at stages $t=2,4,\ldots$. 
\item[ii)]  The same argument guarantees that the active nodes in layer $\beta$ at stage $t=2$ are active 
at the stages $t=4,6, \ldots$, and their replica nodes in $\alpha$ are never reactivated at stages $t=3,5, \ldots$. 
\item[iii)] Statements i) and ii) may  appear to guarantee that, 
when a node has been categorized as inactive, it cannot be reactivated later. 
However, this is not necessarily the case only in Case F. 
This is because it is not ensured that the active nodes in layer $\beta$ at stage $t=2$, the replica nodes of which
in layer $\alpha$ are  inactive (damaged or isolated) at stage $t=1$, form the GC, 
which allows a portion of the inactive nodes at $t=1$ to be reactivated at stage $t=3$. 
\end{itemize} 
Consideration of i)--iii) restricts possible state transitions to 
those depicted in Fig. \ref{transition}. 
This figure indicates that we can terminate the repetition of the percolation at stage $t=3$ in Case Q and at stage $t=4$ in Case F, considering that the percolation processes  converge. 

\subsection{Bipartite graph expression and notation}
\label{section: bipartite}

\begin{figure}[htbp]
\centering
\includegraphics[width=6cm]{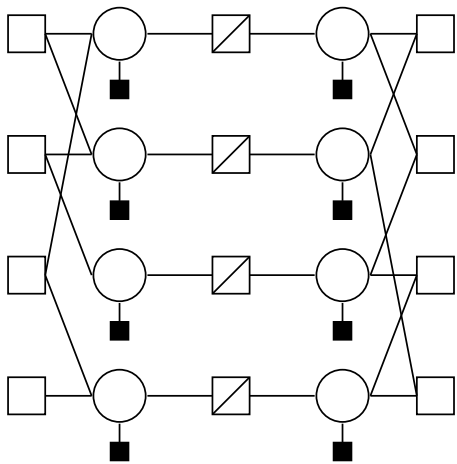}
\caption{Diagram of  antagonistic duplex networks.} 
\label{fig: duplex networks}
\end{figure}
  In Fig. \ref{fig: duplex networks}, we provide the bipartite graph representations of the original networks, which help us  consider the message passing scheme in the failure process graphically.
Each original node is also  a variable node and expressed as a circle.
To indicate 
whether the variable node has initially failed or not without removing it, a function node is  connected to each  variable node, which is depicted as a black plain square in the figure.
For the purpose of  passing messages, we append a function node on each interlink and each intralink, respectively.
A function node that is depicted as a white square with a slash inside  it expresses the role of the interlinks, whereas a function node that is depicted as a white plain square
 represents the role of  the intralink.  
We now introduce the basic notation for antagonistic bipartite networks. 
We denote a variable node in layer $\alpha$ by $\iA$. 
The variable node $\iA$ is directly connected with the set of function nodes,  
which is denoted by $\partial \iA$.
We denote a function node on each intralink in layer $\alpha$ by $\aA$,
and we denote a function node on interlinks by $p$.
The function node $\aA$ is directly connected with two variable nodes, denoted as  $\partial\aA$.

\subsection{Statistical expression of graph topologies}
\label{section: topologies}
One of the most fundamental topologies of network (layer) is  degree distribution, which is defined as the probability that a randomly chosen node has degree $\kA$, denoted by $\pA(\kA)$. 
We also provide $\rA (\kA)$, which denotes the degree distribution of a link computed  as the probability that one terminal node of a randomly chosen link  has  degree $\kA$. We describe  $\rA (\kA)$ using $\pA(\kA)$:
\begin{eqnarray}
\rA(\kA)=\frac{\kA \pA(\kA)}{\sum_\lA \lA \pA(\lA)}.\hspace{2mm}
\label{rA1}
\end{eqnarray}
Related to this, 
the intralayer joint degree distribution ({\it intralayer degree-degree correlations})  is defined as the probability that,  given  an intralink is randomly chosen, one terminal node has degree $\kA$ and the  second terminal node has degree $\lA$, which is denoted by  $\rA(\kA,\lA)$  and is described using $\rA (\kA)$,
\begin{eqnarray} 
\rA(\kA)=\sum_{\lA}\rA(\kA,\lA),\hspace{2mm}
\label{rA2}
\end{eqnarray}
Related to the intralayer degree-degree correlations, the intra-joint degree distribution ({\it intralayer degree-degree correlations}) is described as
\begin{eqnarray}
\rA(\kA|\lA)=\frac{\rA(\kA,\lA)}{\rA(\lA)},\vspace{6mm}
\label{rA3}
\end{eqnarray}
The inter-joint degree distribution  
is denoted by $P(\kA,\kB)$, defined as the probability that the degrees of  a randomly chosen node-pair 
are $\kA$ and $\kB$, and named {\it interlayer degree correlations}.
The relationship between $\pA(\kA)$ and $P(\kA,\kB)$  is
\begin{eqnarray}
\pA (\kA )=\sum_{\kB} P(\kA,\kB),
\label{pA}
\end{eqnarray}
Related to interlayer joint degree distribution, the inter-conditional distribution is described as
\begin{eqnarray}
\PA(\kA|\kB)&=&\frac{\PA(\kA,\kB)}{\sum_{\kB}\PA(\kA,\kB)}, 
\label{PA}
\end{eqnarray}
 which is defined as the probability of  a node having degree $\kA$,  given that the degree of its replica node  is $\kB$.
 
Exchanging $\alpha$ with $\beta$ in Eqs. (\ref{rA1})-(\ref{PA}),   we define $\rB(\kB)$, $\rB(\kB|\lB)$, $\pB (\kB)$, and $\PB(\kB|\kA)$, respectively.
 Using the definition of $\PB(\kB|\kA)$, the intralayer conditional distributions of  {\it node pairs} are obtained 
\begin{eqnarray}
\rA(\kA,\kB|\lA,\lB)&=&\PB(\kB|\kA)\rA(\kA|\lA), 
\label{pair_conditionalA}
\end{eqnarray}
the definition of which is the probability that a randomly chosen node pair 
 having the degree ($\lA$, $\lB$)  is connected with another node pair having the degree ($\kA$, $\kB$),  given an intralink in layer $\alpha$.

\section{Theoretical framework}
\label{section: framework}
The aim of this section is to develop a framework for analyzing the robustness of antagonistic networks based on the cavity method.
In preparation for evaluating the GC size from the macroscopic viewpoint, we examine the message flow for a single instance from the microscopic viewpoint in Sec. \ref{section: micro}; 
Sec. \ref{section: micro1,2} provides the message flow at stages $t=1$ and $t=2$,  and 
Sec. \ref{section: micro (Case Q)}  and Sec. \ref{section: micro (Case F)} describe 
how the flow behaves for $t \ge 3$ in Cases Q and F, respectively. 
In Sec. \ref{section: cross link}, we define macroscopic intralayer messages using microscopic ones. 
With the aid of local tree approximation and self averaging properties of random network, we extend our formulation to the macroscopic level in Sec. \ref{section: macro};
Sec. \ref{section: macro1,2} provide the macroscopic message flow at stages $t=1$ and $t=2$,  and 
Sec. \ref{section: macro (Case Q)}  and Sec. \ref{section: macro (Case F)} describe 
how the macroscopic flow behaves for $t \ge 3$ in Cases Q and F, respectively. 

\subsection{Message flow from the microscopic viewpoint}
\label{section: micro} 
\subsubsection{First and second stage}
\label{section: micro1,2}
We define a binary variable $\psi_{\iA}$, which is set at 0 or 1 depending on whether or not the node suffers from the initial damage, respectively, and  assign it to another variable, named the activity index $s^{t=1}_{\iA}$ of $\iA$ at stage $t=1$, 
\begin{eqnarray}
s^{t=1}_{\iA}\equiv\psi_{\iA}.
\label{s1}
\end{eqnarray}
Note that  the total fraction of the active nodes at the initial condition, namely, $\sum_{i^{}_{\alpha}} \delta\left(\psi_{i^{}_{\alpha}}=1\right)/N$, is handled as a survival ratio in Sec. \ref{section: macro}.
 To examine the first stage in layer  $\alpha$, we apply the cavity method 
 presented in \cite{Mezard} for the given set of $\{s^{1}_{\iA}\}$,  
 which yields a set of self-consistent equations
\begin{eqnarray}
m^{t=1}_{\aA \to \iA} &=& m^{t=1}_{\jA \to \aA} \quad \left (\partial \aA =\{\iA , \jA \} \right ), \label{hstep_A} \\
m^{t=1}_{\iA \to \aA} &=&  1-  s^{t=1}_{\iA}+ s^{t=1}_{\iA}\prod_{{\bA} \in \partial \iA\backslash \aA} m^{t=1}_{\bA \to \iA}. \label{vstep_A}
\end{eqnarray}
Here, $m_{i \to a}^t \in \{0,1\}$ in general denotes the message from variable node $i$ to function node $a$ at the $t$-th stage, 
which takes $0$ when $i$ belongs to a GC in the layer from which node $a$ is removed, and unity, otherwise. 
The message $m_{a \to i}^t \in \{0, 1\}$, on the other hand, 
conveys $0$ from function node $a$ to variable node $i$ at the $t$-th stage
when at least one $j \in \partial a \backslash i$ belongs to the GC, and unity, otherwise. 
Using  the solution of Eqs. (\ref{hstep_A}) and (\ref{vstep_A}), we derive the indices of the GC and the size of the  GC in layer $\alpha$ at stage $t=1$ as
\begin{eqnarray}
\sigma^{t=1}_{\alpha}=\sum_{\iA} \sigma^{t=1}_{\iA}=\sum_{\iA} s^{t=1}_{\iA} \left (1-\hspace{-2mm}\prod_{\aA \in \partial \iA} m^{t=1}_{\aA \to \iA}\right )\hspace{-1mm},
\label{gstep_A1}
\end{eqnarray}
which also provides  a message from  ${\iA}$ to the function node $p$ on an interlink at stage $t=1$ as
\begin{eqnarray}
m^{t=1}_{\iA \to p}&=&\sigma^{t=1}_{\iA}=s^{t=1}_{\iA} \left (1-\prod_{\aA \in \partial \iA} m^{t=1}_{\aA \to \iA}\right ). \label{i->p:A1}
\end{eqnarray}
Because of the antagonistic nature of the interlinks, the inverted value of Eq. (\ref{i->p:A1}) is propagated 
from the function node $p$ to the replica node $\iB$ of layer $\beta$ after stage $t=1$ as
\begin{eqnarray}
1 - m^{t=1}_{\iA \to p}&=& 1 - s^{t=1}_{\iA} \left (1-\prod_{\aA \in \partial \iA}m^{t=1}_{\aA \to \iA}\right )\cr
&=& m^{t=2}_{p\to\iB} .
\label{iA->iB}
\end{eqnarray}

The second stage ($t=2$) is considered the initial step for layer $\beta$.
In contrast to  the first stage at layer $\alpha$, a particular set of $\psi_{\iB}$ is not involved
(in other words $\psi_{\iB}=1$), 
because the nodes in layer $\beta$ are free from the initial damage and influenced only by the activity pattern of layer $\alpha$ provided by the step at stage $t=1$.
The activity index of ${\iB}$ at  the start of stage $t=2$ is
\begin{eqnarray}
s^{t=2}_{\iB}  
 = m^{t=2}_{p\to \iB} = 1-\sone\left (1-\prod_{\aA \in \partial \iA} m^{t=1}_{\aA \to \iA}\right )\hspace{-1mm}. 
\label{siB}
\end{eqnarray}
Note that $s^{t=2}_{\iB} =1$ holds, if  either $\psi_{\iA}=0$ or  $\psi_{\iA}=1$ and $\prod_{\aA \in \partial \iA} m^{t=1}_{\aA \to \iA}=0$ are satisfied. 
 Given $\{s^{t=2}_{\iB}\}$, the cavity method provides
  the  self-consistent equations
\begin{eqnarray}
m^{t=2}_{\aB \to \iB} &=& m^{t=2}_{\jB \to \aB} \quad \left (\partial \aB =\{\iB ,\jB \} \right ), \label{hstep_B} \\
m^{t=2}_{\iB \to\aB} &=&  1-  s^{t=2}_{\iB}  + s^{t=2}_{\iB}  \prod_{\bB \in \partial \iB
\backslash \aB} m^{t=2}_{\bB \to \iB}. \label{vstep_B}
\end{eqnarray}
The solution of Eqs. (\ref{hstep_B}) and (\ref{vstep_B}) provides
the GC size of layer $\beta$ at stage $t=2$ as
\begin{eqnarray}
\sigma^{t=2}_{\beta} =\sum_{\iB} \sigma^{t=2}_{\iB}=\sum_{\iB} s^{t=2}_{\iB} \left (1-\hspace{-2mm}\prod_{\aB \in \partial \iB} m^{t=2}_{\aB \to \iB}\right )\hspace{-1mm}. \label{gstep_B}
\end{eqnarray}
The solution also provides the messages from ${\iB}$ to $p$ and the message from ${p}$ to $\iA$ as
\begin{eqnarray}
m^{t=2}_{\iB \to p}&=&s^{t=2}_{\iB} \left (1-\prod_{\aB \in \partial \iB} m^{t=2}_{\aB \to \iB}\right ).\label{i->p:B}\\
m^{t=3}_{p\to \iA} &=&1 - m^{t=2}_{\iB \to p}.\label{iB->iA}
\end{eqnarray}
At the third and further stages, nodes in layer $\alpha$ receive the inter-messages  that are represented as Eq. (\ref{iB->iA}). 

\subsubsection{Third and further stages in Case Q}
\label{section: micro (Case Q)}
As already discussed  in Sec. \ref{section: process},  we obtain the final robustness of layer $\alpha$ and layer $\beta$, which is the robustness of layer $\alpha$ at the  first stage (Eq. (\ref{gstep_A1})) and that of  layer $\beta$ at the second  stage in  (Eq. (\ref{gstep_B})), respectively.
\begin{eqnarray}
\sigma^{2t\rq{}+1}_{\iA}  = \sigma^{t=1}_{\iA},  
\sigma^{2t\rq{}}_{\iB}  =\sigma^{t=2}_{\iB} {\rm (Q)}.
\label{sigmaQfinal}
\end{eqnarray}

\subsubsection{Third and further stages in Case F}
\label{section: micro (Case F)}
In Case F, nodes in layer $\alpha$ at stage $t=3$ are influenced by only interlayer messages and are free from the initial activity indexes, which provides the activity index of ${\iA}$ as 
\begin{eqnarray}
s^{t=3}_{\iA} {\rm (F)} &=& m^{t=3}_{p\to \iA} \cr
&=& 1-s^{t=2}_{\iB}\left(1-\prod_{\aB \in \partial \iB} m^{t=2}_{\aB \to \iB}\right)
\label{s_iA3 (F)}
\end{eqnarray}
Substituting $s^{t=3}_{\iA}{\rm (F)} $ for $s^{t=1}_{\iA}$
 in Eqs. (\ref{hstep_A}) and (\ref{vstep_A}), 
 
\begin{eqnarray}
m^{t=3}_{\aA \to \iA} &=& m^{t=3}_{\jA \to \aA} \quad \left (\partial \aA =\{\iA , \jA \} \right ), \label{hstep_A3} \\
m^{t=3}_{\iA \to \aA} &=& 1- s^{t=3}_{\iA}{\rm (F)} + s^{t=3}_{\iA}{\rm (F)}\prod_{{\bA} \in \partial \iA\backslash \aA} m^{t=3}_{\bA \to \iA},\cr
&&
\label{vstep_A3}
\end{eqnarray}
 we obtain the solution $m^{t=3}_{\aA \to \iA}$, 
 which derives the size of the GC  in layer $\alpha$  as
\begin{eqnarray}
\sigma^{t=3}_{\alpha}&=&\sum_{\iA} \sigma^{t=3}_{\iA}=\sum_{\iA} s^{t=3}_{\iA}{\rm (F)} \left (1-\hspace{-2mm}\prod_{\aA \in \partial \iA} m^{t=3}_{\aA \to \iA}\right )\cr
&& 
\label{gstep_A3F}
\end{eqnarray}

 We describe $s^{t=4}_{\iA} {\rm (F)}$ that denotes the message  through the interlink, which each of $\iB$ receives at  stage $t=4$, as 
\begin{eqnarray}
s^{t=4}_{\iA} {\rm (F)} &=& m^{t=4}_{p\to \iB}=m^{t=3}_{\iA\to p}=1- \sigma^{t=3}_{\iA}\cr
&=& 1-s^{t=3}_{\iB}{\rm (F)}\left(1-\prod_{\aB \in \partial \iB} m^{t=3}_{\aB \to \iB}\right).
\label{s_iB4 (F)}
\end{eqnarray}
As discussed  in Sec. \ref{section: process}, the percolation result at stage $t=4$  becomes identical to that at stage $t=2$ in Case F. 
Therefore, we can determine the  indices  of each node in each network as
\begin{eqnarray}
\sigma^{2t\rq{}+1}_{\iA}  = \sigma^{t=3}_{\iA},  
\sigma^{2t\rq{}}_{\iB}  =\sigma^{t=2}_{\iB} {\rm (F)}.
\label{sigmaFfinal}
\end{eqnarray}
Consequently, we can terminate the repetition of the percolation at stage $t=2$
considering the networks converged (Fig. \ref{transition} (a)). 

\begin{figure}[htbp]
\begin{center}
\includegraphics[width=8cm]{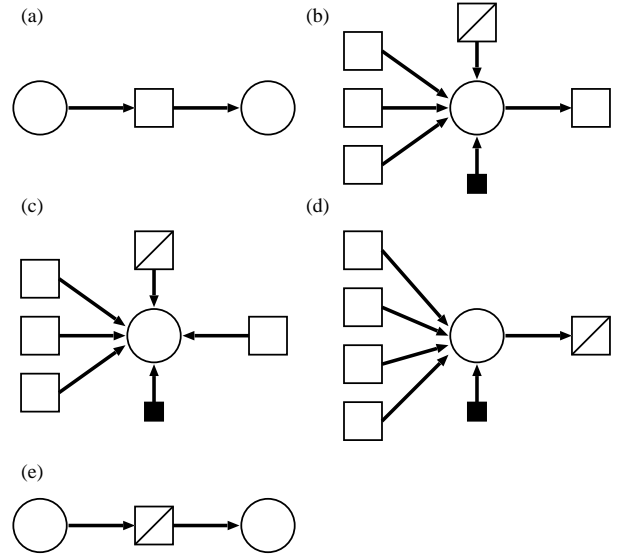}
\end{center}
\caption{Diagram of message passing.} 
\begin{enumerate}
\item The message flow passing a function node in a layer, which corresponds to Eq. (\ref{hstep_A}), Eq. (\ref{hstep_B}) and  Eq. (\ref{hstep_A3}).
\item The message flow passing a variable node in a layer,  which corresponds to Eq. (\ref{vstep_A}), Eq. (\ref{vstep_B}) and  Eq. (\ref{vstep_A3}). 
\item The message flow for computing the size of the GC,  which corresponds to Eq. (\ref{gstep_A1}), Eq. (\ref{gstep_B}), and Eq. (\ref{gstep_A3F}). 
\item The message flow from a variable node in a layer to a function node on an interlink, which corresponds to Eq. (\ref{i->p:A1}) and Eq. (\ref{i->p:B}). 
\item The message flow from  a function node on an interlink to a variable node in the layer,  which corresponds to Eq. (\ref{iA->iB}) or Eq. (\ref{iB->iA}). 
\end{enumerate}
\label{fig: message flow}
\end{figure}

The local message flows are categorized  with the aid of the bipartite graph expression  that is introduced in Sec. \ref{section: bipartite} (See Fig. \ref{fig: message flow}).
\renewcommand{\labelenumi}{\theenumi}
\renewcommand{\theenumi}{(\alph{enumi})}
\subsection{Cross link from the microscopic viewpoint to the macroscopic one}
\label{section: cross link}
We first focus on a node pair  $\iA$ and $\iB$, the degrees of which are $\lA$ and $\lB$, respectively. The node  $\iA$  is initially attached $s_{\iA}^{t=1}$, which is described as  Eq. (\ref{s1}).
Using $s_{\iA}^{t=1}$, we evaluate $q_{\lA\lB}^{{\alpha},t=1}$,  which denotes the fraction of the {\it set} of node pairs, the degrees of which are  $\lA$ and $\lB$, taking the value of unity at  the first stage:
\begin{eqnarray}
q^{\alpha, {t=1}}_{\lA,\lB}&=&
\frac{\sum_{\iA} \delta\left(|\partial \iA|=\lA\right) \delta\left(|\partial \iB|=\lB\right)s^{\alpha, {t=1}}_{\iA}}
{\sum_{\iA} \delta\left(|\partial \iA|=\lA\right)\delta\left(|\partial \iB|=\lB\right)}.\cr
&&
\label{qAdef}
\end{eqnarray}
In the case of layer $\beta$, we denote the relevant active probability by 
\begin{eqnarray}
q^{\beta, t=2}_{\lA,\lB}&=&
\frac{\sum_{\iB} \delta\left(|\partial \iA|=\lA\right) \delta\left(|\partial \iB|=\lB\right)s^{\beta, t=2}_{\iB}}
{\sum_{\iB} \delta\left(|\partial \iA|=\lA\right)\delta\left(|\partial \iB|=\lB\right)}.
\label{qBdef}
\end{eqnarray}
 As in  Eq. (\ref{qAdef}), we  implicitly define $q^{\alpha, {t=3}}_{\lA,\lB}$  using $s^{\alpha, {t=3}}_{\iA}$, which leads to Eq. (\ref{qA3}). 
 We  compute the fraction of  message $m^{t=1}_{\jA \to \aA}=m^{t=1}_{\aA \to \iA} $ taking unity, which is characterized by  the degree of $\iA$ and its replica node $\iB$,  solving the relevant cavity equations ({\it e.g.}, Eq. (\ref{macro_A1})) iteratively. We  denote the macroscopic message by $I^{\alpha, t=1}_{\lA,\lB}$: 
\begin{eqnarray}
I^{\alpha, {t=1}}_{\lA,\lB} &=&
\frac{\sum_{\iA} \delta\left(|\partial \iA|=\lA\right) \sum_{\aA \in \partial \iA} m^{\alpha, t=1}_{\aA\to \iA} }
{\lA  \sum_{\iA} \delta\left(|\partial \iA|=\lA\right)\delta\left(|\partial \iB|=\lB\right)}.
\label{IAdef}
\end{eqnarray}
Similarly, in layer $\beta$,
\begin{eqnarray}
I^{\beta, t=2}_{\lA,\lB} &=&
\frac{\sum_{\iB} \delta\left(|\partial \iB|=\lB\right) \sum_{\aB \in \partial \iB} m^{\beta, t=2}_{\aB \to\iB} }
{\lB  \sum_{\iB} \delta\left(|\partial \iA|=\lA\right)\delta\left(|\partial \iB|=\lB\right)}.
\label{IBdef}
\end{eqnarray} 
Note that these macroscopic variables are relevant for network ensembles, and thus, may  not be appropriate for each individual network; in particular, each active label of all the nodes  is strongly correlated with the network connectivity.
 
\subsection{Message flow from the macroscopic viewpoint}
\label{section: macro}
\subsubsection{First and second stage}
\label{section: macro1,2}
 Let us consider the initial stage ($t=1$)  in layer $\alpha$.
First, we set only one parameter  $q$ as the initial  survival probability, that is, 
the fraction of non-failed nodes in layer $\alpha$.
In the case of RFs, we set
\begin{eqnarray}
q^{\alpha,t=1}_{\kA,\kB} =q^{\alpha,t=1}_{\kA}= q, \label{qA1;RF}
\end{eqnarray}
On the other hand, in the case of TAs,  nodes having larger degrees in layer $\alpha$ are preferentially damaged. 
We therefore set 
\begin{eqnarray}
q^{\alpha,t=1}_{\kA,\kB} =q^{\alpha,t=1}_{\kA} \left \{
\begin{array}{ll}
0 & (\kA> \Theta) \cr
\Delta & (\kA= \Theta), \\
1 & (\kA< \Theta) 
\end {array}
\right. \label{qA1;TA} 
\end{eqnarray}
where $\Theta$ and $\Delta$ are uniquely determined so that 
\begin{equation}
q= \sum_{\lB} \left (\Delta P(\Theta, \lB)+\sum_{\lA < \Theta } P(\lA, \lB) \right )
\end{equation}
holds. 
Substituting $\qAone$ in the self-consistent equation
\begin{eqnarray}
\IAllone
&&= \sum_{\kA,\kB} \rA \left( \kkll\right)\cr&&\left(1-\qAone+\qAone \IAkkkmione \right),
\label{macro_A1}
\end{eqnarray}
which corresponds to Eqs. (\ref{hstep_A}) and (\ref{vstep_A}).
 Solving Eq. (\ref{macro_A1}) iteratively, the solution $\IAllone$ is determined,
which offers the fraction of GC in layer $\alpha$ at stage $t=1$ 
\begin{eqnarray}
\hspace{0mm}\mu^{t=1}_{\alpha} 
=  \sum_{\kA,\kB} \Pkk\qAone \left (1-\IAkkkone\right ).  
\label{macro;GCA1}
\end{eqnarray}
Let us consider the second stage in layer $\beta$,
in which $\qBtwo$ denotes  the probability that nodes do not fail at stage $t=2$, the degree of which is $\kB$; their  replica node's degree  is $\kA$.    
Considering the message flow (Eq.(\ref{i->p:A1}), Eq. (\ref{iA->iB}), and Eq. (\ref{siB})),  $\qBtwo$ is  directly calculated from the solution  $\IAllone$ in Eq. (\ref{macro_A1}) as 
\begin{eqnarray}
\qBtwo&=&1-\qAone +\qAone \IAkkkone.
\label{qB2}
\end{eqnarray}
Substituting $\qBtwo$ in the self-consistent equation
\begin{eqnarray}
\IBlltwo&=& \sum_{\kA,\kB} \rB \left( \kkll\right) \cr
&&\left (1 -\qBFtwo+\qBFtwo \IBkkkmitwo \right ), \cr
&&
\label{macro_B}
\end{eqnarray}
 based on Eqs. (\ref{hstep_B}) and (\ref{vstep_B}),  we compute the set of messages $\IBlltwo$,
which offers the fraction of the GC in layer $\beta$ at stage $t=2$ as 
\begin{eqnarray}
\mu^{t=2}_{\beta}
=  \sum_{\kA,\kB} \Pkk\qBtwo \left (1-\IBkkktwo \right ). 
\label{macro;GCB}
\end{eqnarray}
\subsubsection{Third and further stages in Case Q}
\label{section: macro (Case Q)}
In Sec. \ref{section: process}, we already discussed that the final GC in layer $\alpha$ is  identical  to that at stage $t=1$, while the final GC  in layer $\beta$ is identical  to that at stage $t=2$, which provides
\begin{eqnarray}
\mu_{\alpha}=\mu^{t=1}_{\alpha}, \mu_{\beta}=\mu^{t=2}_{\beta}.
\end{eqnarray}
\subsubsection{Third and further stages in Case F}
\label{section: macro (Case F)}
Here, we naively  compute $\qAthree$, the fraction of nodes that are not failed at stage $t=3$, based on Eq. (\ref{s_iA3 (F)})   
 \begin{eqnarray}
\qAthree
&=&1-\qBtwo+\qBtwo \IBkkktwo .
\label{qA3}
\end{eqnarray}
As discussed in Sec. \ref{section: numerical_test},  it is necessary to  evaluate Eq. (\ref{s_iA3 (F)})  in detail.
Substituting $\qAthree$ in the self-consistent equation based on Eqs. (\ref{hstep_A3}) and (\ref{vstep_A3})
\begin{eqnarray}
\IAllthree
&&= \sum_{\kA,\kB} \rA \left( \kkll\right)\cr&&\left(1-\qAthree+\qAthree \IAkkkthree \right),
\label{macro_A3}
\end{eqnarray}
we obtain a solution of $\IAkkthree$, which yields  the fraction of the GC in layer $\alpha$
 \begin{eqnarray}
\mu^{t=3}_{\alpha} 
=  \sum_{\kA,\kB} \Pkk\qAthree \left (1-\IAkkkthree\right ) \cr
&& 
\label{GCA_antagonistic}
\end{eqnarray}
The GC in layer $\beta$ at stage $t=4$ is identical to that at stage $t=2$, which means that the percolation process
is only the repetition of the stage at stage $t=3$ and the stage at stage $t=4$ alternately.
Therefore, the robustness of layer $\alpha$ and layer $\beta$ is evaluated as 
 \begin{eqnarray}
 \mu_{\alpha } \rm (F)=\mu^{t=3}_{\alpha}  {\rm (F)},
 \mu_{\beta} \rm (F)=\mu^{t=2}_{\beta}  {\rm (F)},
\label{GCFfinal}
\end{eqnarray}
respectively.
\section{Numerical test}
\label{section: numerical_test}
\renewcommand{\labelenumi}{\theenumi}
\renewcommand{\theenumi}{(\arabic{enumi})}
\subsection{Procedure}
\label{section: procedure}
We conducted numerical experiments to  confirm the validity of the developed 
method for analyzing the robustness of antagonistic networks. 
Here, the procedure of the numerical experiments is briefly described.
\begin{enumerate}
\item We constructed two random networks (layers) $\alpha$ and $\beta$, the size of each of which was $N=10000$. The degree distribution of each layer was represented by $\PA(4)=\PB(4)=0.5, \PA(6)=\PB(6)=0.5$.  
\item To introduce intralayer degree-degree correlations,  we set a Pearson coefficient in each layer, $C_{\alpha}$ and $C_{\beta}$, respectively. For each layer, we randomly selected two pairs of connected nodes  and rewired the intralinks, employing the algorithm in \cite{Newman2002}. 
\item  To introduce interlayer degree-degree correlations,  we set a Pearson coefficient between layers, $C_{I}$. We rewired the interlinks, reordering the indices of one layer. Note that  it is necessary to suppose that $P(\kA=x, \kB=y)=P(\kA=y, \kB=x)$, because $C_{I}$ does not determine $P(\kA,\kB)$ uniquely. 
\item For the degree-correlated networks, we applied the Monte Carlo simulation described below.  
Setting an initial survival probability $q$, we chose initially failed nodes randomly depending on the type of failures (RFs or TAs).   Failed nodes cause networks to decompose into connected components,  each of which is detected using the algorithm in  \cite{HK, Al-Futaisi}.  
Note that we modified the open MATLAB code in \cite{Al-Futaisi} very slightly in the part of  \lq\lq{}case 4c ii\rq\rq{}, tracing the nest of \lq\lq{}NodeLP(N)\rq\rq{} {\it sufficiently}  to reach its source such  that  \lq\lq{}NodeLPmin\rq\rq{} should be put the least cluster label. 
We terminated  the single instance if  each active label of all nodes in layer $\alpha$   accorded with that at the last stage in a one-to-one manner.  
Similar procedures were tested 50 times at each initial survival probability, $q$.
\end{enumerate}
\subsection{Methodological accuracy}
\subsubsection{Case Q}
\begin{figure}[h]
\begin{center}
\includegraphics[width=8cm,height=8cm]{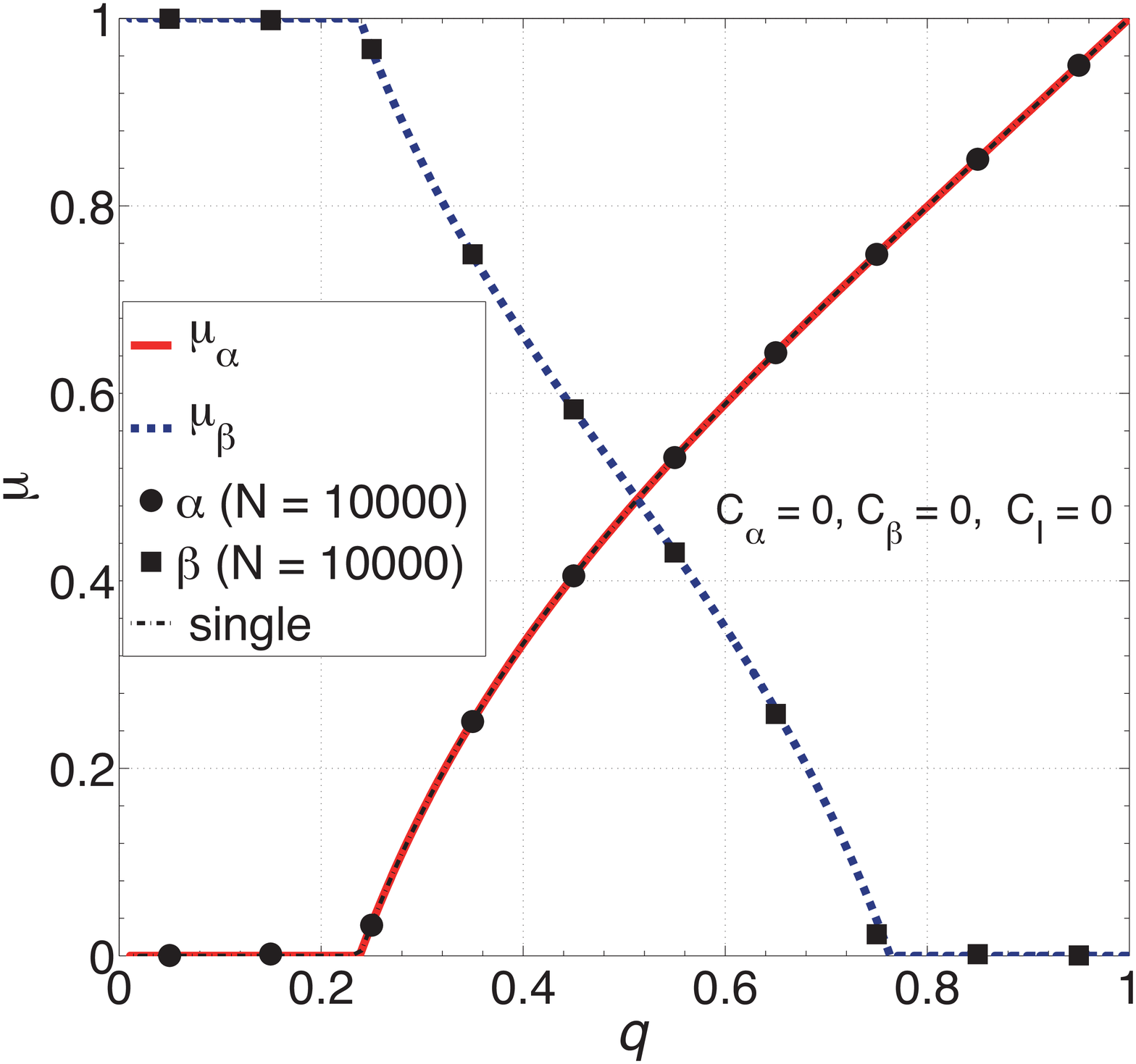}
\end{center}
\caption{
(Color on-line)   Analytical results of the robustness of layer $\alpha$  versus $q$ for the case where  antagonistic networks suffer from RFs and the setting is Case Q.  
Examples of robustness of layer $\alpha$ at stage $t=1$ are also plotted, which are the results of failure processes that are completed in the single layer $\alpha$.
Each dot  is averaged 50 times, produced by numerical experiments. 
\label{fig. (Q)}}
\end{figure}
\begin{figure}[h]
\includegraphics[width=8cm,height=8cm]{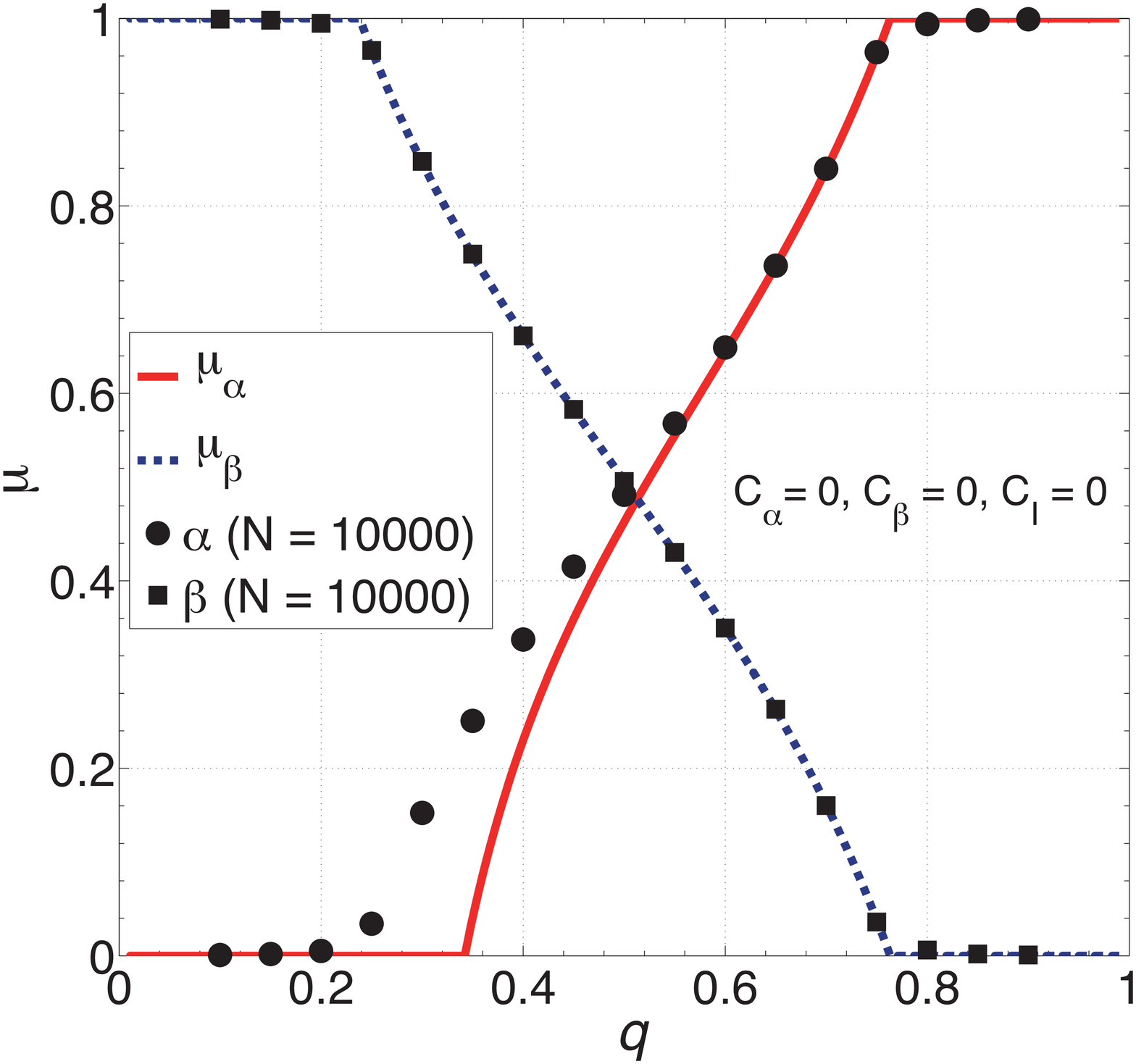}
 \caption{Analytical results of the robustness of layer $\alpha$  versus $q$ for the case where the antagonistic networks suffer from RFs and the setting is Case F.  Each dot  is averaged 50 times, produced by numerical experiments. 
 \label{fig:free1}}
\end{figure}

Fig. \ref{fig. (Q)} shows a comparison of the theoretical prediction obtained 
for the analysis  and the experimental results, 
which exhibits an excellent consistency. 
In particular, antagonistic interlinks do not affect the robustness in layer $
\alpha$, which is the GC at stage $t=1$. 
\subsubsection{Case F}
\label{subsection: result Free}
 Fig. \ref{fig:free1} shows a comparison  of the theoretical predictions of   the robustness of layers $\alpha$  and $\beta$  and the numerical results in the condition of Case F.
Experimental data for  layer $\beta$ exhibit excellent accordance with the theoretical predictions.  
However, with regard to the robustness in layer $\alpha$, there exist significant 
discrepancies between the theoretical predictions and the numerical results. 

\section{Accuracy improvement}
\label{section: improvement}
The results in Sec. \ref{section: numerical_test} indicate that 
there are significant discrepancies in the GC size evaluation between theory and experiment for Case F. 
The purpose of this section is to resolve this inconsistency. 
In Sec. \ref{section: cause}, we examine the cause of the discrepancies. 
To improve the accuracy of the theoretical evaluation, we derive alternative expressions for the  microscopic Eqs. (\ref{s_iA3 (F)})-(\ref{gstep_A3F}) and corresponding macroscopic Eqs. (\ref{qA3})-(\ref{GCA_antagonistic}).  
Consequently,  we  introduce a heuristic treatment in the microscopic level  in Sec. \ref{section: heuristic(micro)} and  extend it to  derive macroscopic expressions in Sec. \ref{section: heuristic(macro)}, the utility of which is verified by numerical experiments. 

\subsection{Clarifying the cause of the discrepancy}    
\label{section: cause}
\begin{figure}[htbp]
\begin{center}
\includegraphics[width=7cm]{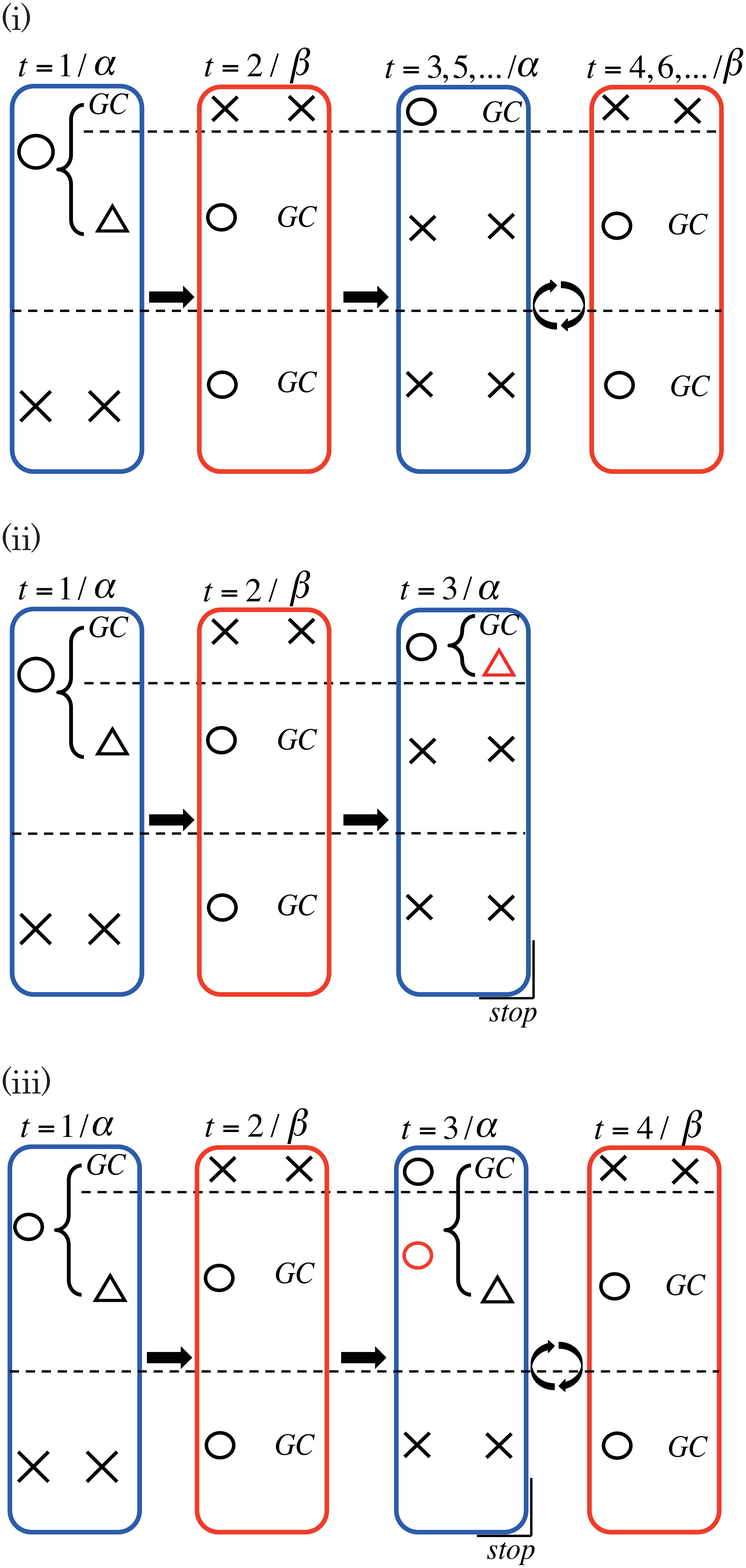}
\end{center}
\caption{
(Color on-line) 
Possible transitions for Case F from  (i) a microscopic,  (ii) a naive macroscopic, and  (iii) a modified macroscopic viewpoint.  Note that,  for simplicity,  we considered the situations where   the  number of active nodes  that were isolated from the GC  in layer $\beta$ was {\it almost negligible}.
The meaning of each symbol is similar to that in Fig. \ref{transition}.  
The red triangle in (ii) represents the nodes that are not judged constituents of the GC, which causes the discrepancies in Fig. \ref{fig:free1}.
The red circle in (iii) represents the nodes that are actually inactive according to the rule of antagonistic interlinks (see also (i)), but regarded active by dropping  the self-feedback term from Eq. (\ref{s3F1}) as Eq. (\ref{s3mod}).
\label{transition_free_thresholds}
}
\end{figure}
The causes  of the above discrepancies between theory and experiment
lie in  the transformation  from microscopic variables to macroscopic ones at stage $t=3$.  
To reach this resolution, we dissect the GC at stage $t=3$, which is constitutively heterogeneous and  divided into three subsets depending on the history of nodes.
\renewcommand{\labelenumi}{\theenumi}
\renewcommand{\theenumi}{(\Roman{enumi})}
\begin{enumerate}
\item Nodes that belonged to the GC at stage $t=1$. They necessarily belonged to the GC at stage $t=3$, which  implies  that  there existed strong correlations between $s^{t=3}_{\iA}$ and  $\Mthree$ and thus $\qAthree$ and  $\IAkkthree$. Note that the last statement holds if and only if node $\iA$ is classified as this class.

\item Nodes that failed at stage $t=1$.  They generally belonged to  the GC  by themselves.
\item Nodes that belonged to  one of the small components at stage $t=1$. They  belonged to the GC at $t=3$ with the aid of  node(s) of (II).
\end{enumerate}
Because of these heterogeneity due to the hysteresis, the assumption that   nodes of the same degree are statistically equivalent   does not hold from the macroscopic viewpoint at stage $t=3$.  Therefore  Eq. (\ref{GCA_antagonistic}),  in which  $\qAthree$ and  $\IAkkthree$ are  independent of each other, underestimate the GC size (See Fig. \ref{transition_free_thresholds} (i) and (ii)).

In order not  to treat the  heterogeneity argued above,  we drop the term of the past intralayer messages  to  obscure  (or {\it  encapsulate}) the  connectivity in the relevant layer in Eq. (\ref{s_iA3 (F)}) .
Note that  unlike interdependent networks,
this treatment has a possibility  of considering that some of  failed nodes belong to the GC in layer $\alpha$  at stage $t=3$ in antagonistic networks, which depends on the percolation result of layer $\beta$ at stage $t=2$ (See Fig. \ref{transitionmod}).
To compensate this inconsistency, we deduct them using the original interlayer message to derive the GC size at microscopic level.   
    
\subsection{Heuristic}
\label{section: heuristic(micro)}
\begin{figure}[t]
\begin{center}
\includegraphics[width=7cm]{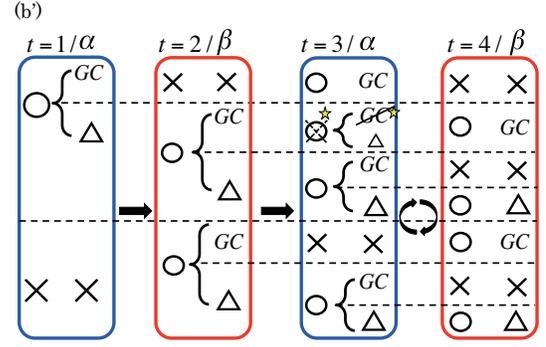}
\end{center}
\caption{
(Color on-line) 
 Possible transitions from a modified macroscopic viewpoint for Case F in the situations where the  number of active nodes  that are isolated from the GC  in layer $\beta$ is {\it not negligible}.
Because some initially destroyed nodes in layer $\alpha$ are revived at stage $t=3$ because of antagonistic interlinks,  some nodes that are treated as active by encapsulation  involuntarily belong to the GC at stage $t=3$, which allows the impossible state transitions that are highlighted by stars.
To compensate this inconvenience, we modified the evaluation of the GC  size as  Eq. (\ref{g3mod}), highlighted by  a small diagonal line.
The meaning of each symbol is the same as in Fig. \ref{transition}.  
\label{transitionmod}
}
\end{figure}

To remain our model analytically tractable, we define the provisional active variables $s^{\ast t=3}_{\iA}$ instead of $s^{t=3}_{\iA} {\rm (F)}$, such that  they do not include messages in layer $\alpha$.
We first describe  Eq. (\ref{s_iA3 (F)}) in detail  and show  that  $s^{t=3}_{\iA} {\rm (F)}$ includes the past  intralayer messages in layer $\alpha$.
\begin{eqnarray}
s^{t=3}_{\iA} {\rm (F)}
&=& \sone+\left( 1-\sone\right)\Mtwo\cr
&-&\sone\Mone\left(1-\Mtwo\right),\cr
&&
\label{s3F1}
\end{eqnarray}  
It is clear that $s^{t=3}_{\iA} {\rm (F)}$ is already influenced by the connectivity of layer $\alpha$, because it includes the term $\Mone$, which is indeed $\Mthree$  itself in the case that node $\iA$ has the history (I) (see Table \ref{table1}).
\begin{table}
\begin{tabular}{|c|c|c||c|||c||c|}\hline
$\sone$& 
$\mone$&
$\mthree$&
$\mthreec$&  
$\sone\mone\mthree$ &
$\sone\mone\mthreec$ 
 \\ \hline
 1& 1 & 1 & 1 & 1& 1\\ \hline
 1& 1 & 0 & 0 & 0& 0\\ \hline
 1& 0 & 0 & 0 & 0& 0 \\ \hline
 1& 0 & 1 & 1 &  $\phi$& $\phi$ \\ \hline
\end{tabular}
\caption{All possible cases of $\Mone$, $\Mthree$, or  $\Mcthree$,   denoted by $\mone$, $\mthree$, and $\mthreec$, respectively. 
$\phi$ denotes the unrealizable case in which $\Mone$ vanishes and $\Mthree$ takes a unity at the same time.
\label{table1}}
\end{table}
Supposing the term $\Mone=0$ in Eq. (\ref{s3F1}), we define the provisional active variables,
\begin{eqnarray}
  s^{\ast t=3}_{\iA} {\rm (F)}&\equiv&\sone+\left( 1-\sone\right)\Mtwo, 
\label{s3mod}  
\end{eqnarray}  
Substituting $s^{\ast t=3}_{\iA}$ for  $s^{t=3}_{\iA} {\rm (F)}$ in Eqs. (\ref{hstep_A3}) and (\ref{vstep_A3}), we compute the provisional message $m^{\ast t=3}_{\aA \to \iA}$.
%
Employing $s^{\ast t=3}_{\iA}$ and $m^{\ast t=3}_{\aA \to \iA}$ in Eq. (\ref{gstep_A3F}),
we can compute the provisional GC, 
\begin{eqnarray}
\sigma^{\ast t=3}_{\iA}
&\equiv&s^{\ast t=3}_{\iA}{\rm (F)}\left(1-\Mcthree\right),
\end{eqnarray}
the sum of which is larger than actual GC size  in particular  when the number of isolated nodes in layer $\beta$ is not negligible (See Fig. \ref{transitionmod}).

Our idea is to replace  only intralayer messages in  Eq. (\ref{gstep_A3F}):
employing $m^{\ast t=3}_{\aA \to \iA}$  instead of  $m^{t=3}_{\aA \to \iA}$,
 we approximately describe the GC label of each node at stage $t=3$ (Eq. (\ref{gstep_A3F}))  in detail:   
\begin{eqnarray}
\sigma^{t=3}_{\iA}  {\rm (F)}
&\approx&s^{\ast t=3}_{\iA}{\rm (F)}\left(1- \Mcthree\right)\cr
&-&\sone\Mone\cr
&\cdot&\left(1-\Mtwo\right)\hspace{-1mm}\left(1-\hspace{-1mm}\Mcthree\right), \cr
&&
\label{trueG}
\end{eqnarray}
where  it is possible to replace $\Mone\Mthree$ with $\Mthree$, which is due to the correlations between  the product of intralayer messages at stage $t=1$ and that at stage $t=3$
(see Table \ref{table1}).  
Therefore, we  renew Eq. (\ref{trueG}) as
\begin{eqnarray}
\sigma^{t=3}_{\iA} {\rm (F)}
&\approx&\sigma^{\ast t=3}_{\iA}
-\sone\left(1-\Mtwo\right)\cr
&\cdot& \left(\Mone-\Mcthree\right),
\label{g3mod}
\end{eqnarray}
 where the second term of Eq. (\ref{g3mod})  corresponds to  deducting the nodes that  incorrectly belong to the GC.
 

\subsection{Macroscopic evaluation and numerical validation}
\label{section: heuristic(macro)}
 For the purposes of macroscopic analysis,  we denote the ratio of active nodes at $t=3$ based on Eq. (\ref{s3mod})
\begin{eqnarray}
\qAastthree&=&\qAone+\left(1-\qAone\right)\IBkkktwo,
\label{qA3new}
\end{eqnarray} 
Substituting $\qAastthree$ in the self-consistent equation
\begin{eqnarray}
\IAastllthree
&&= \sum_{\kA,\kB} \rA \left( \kkll\right)\cr
&&\cdot\left(1-\qAastthree+\qAastthree \IAastkkkthree \right)\hspace{-1mm},
\label{macro_A3new}
\end{eqnarray}
we compute the set of messages $\IAastllthree$, which  yields the fraction of the GC
 \begin{eqnarray}
 \mu^{\star t=3}_{\alpha {\rm (F)}}& 
= &\sum_{\kA,\kB} \Pkk \qAastthree\left (1-\IAastkkkthree\right ) \cr
&-& \qAone \left(1-\IBkkktwo\right)\cr
&\cdot&\left(\IAkkkone-\IAastkkkthree\right),
\label{mu3}
 \end{eqnarray}
based on Eq. (\ref{g3mod}). 
\begin{figure}[tbp]
\includegraphics[width=8cm,height=8cm]{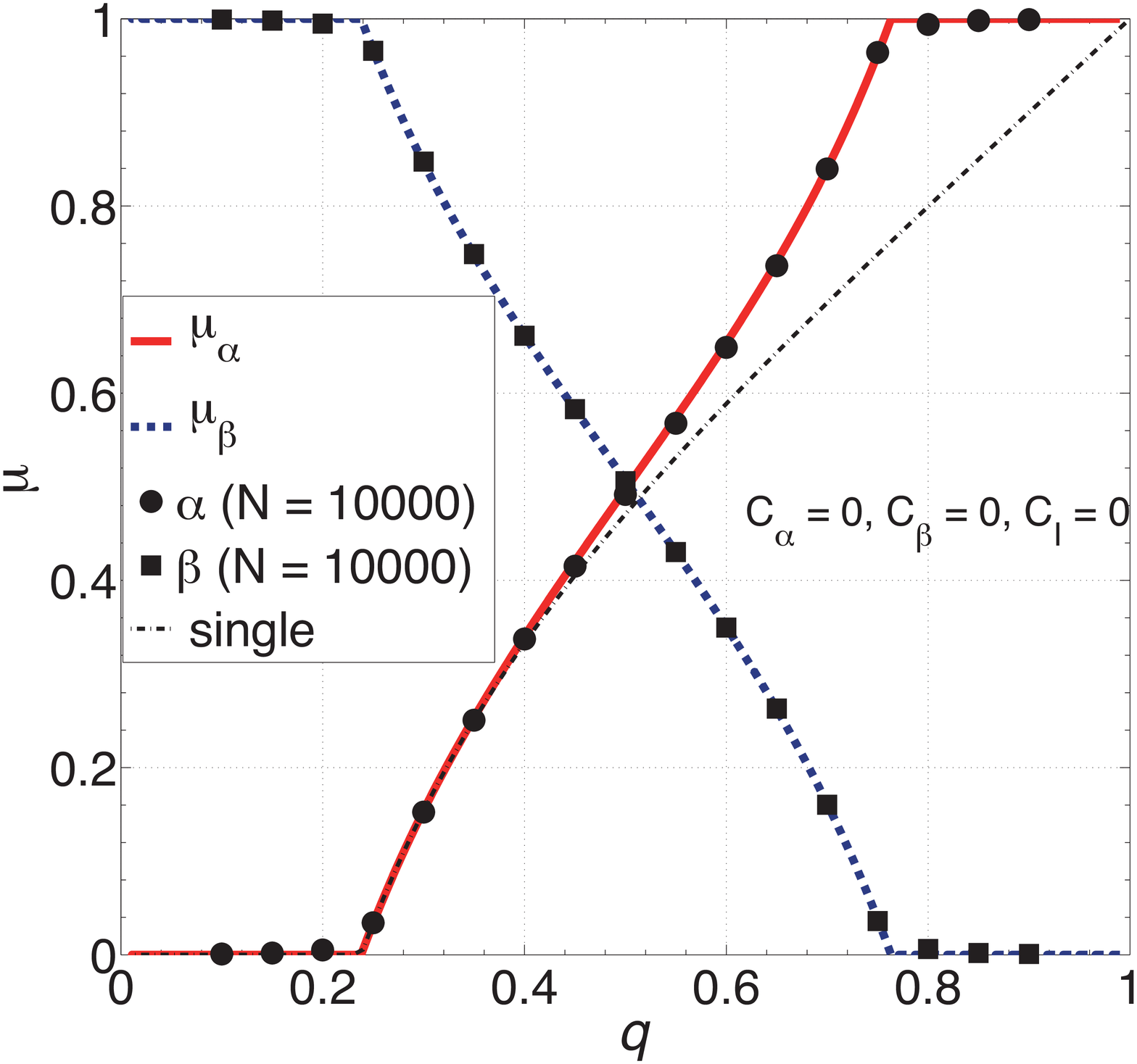}
 \caption{ 
 (Color on-line) Modified analytical results of the robustness of layer $\alpha$  versus $q$ for the case where antagonistic networks suffer from RFs and the setting is Case F. Examples of robustness of layer $\alpha$ at stage $t=1$ are also plotted, which are the results of failure processes that are completed in the single layer $\alpha$.
Each dot  is averaged 50 times, produced by numerical experiments. 
 \label{fig:free2}}
\end{figure}

Fig. \ref{fig:free2} shows the  size of the GCs in layer $\alpha$ predicted by Eq. (\ref{mu3}) in  the case where no degree-degree correlations exist, which excellently accords with the experimental data. As long as we examined, similar accuracy was also achieved for the other parameter sets. 


\section{Result}
\label{section: result}
\subsection{Influence of degree-degree correlaions}
\label{section: deg-correlation}
\begin{figure}[tp]
\begin{center}
\includegraphics[width=8cm,height=7cm]{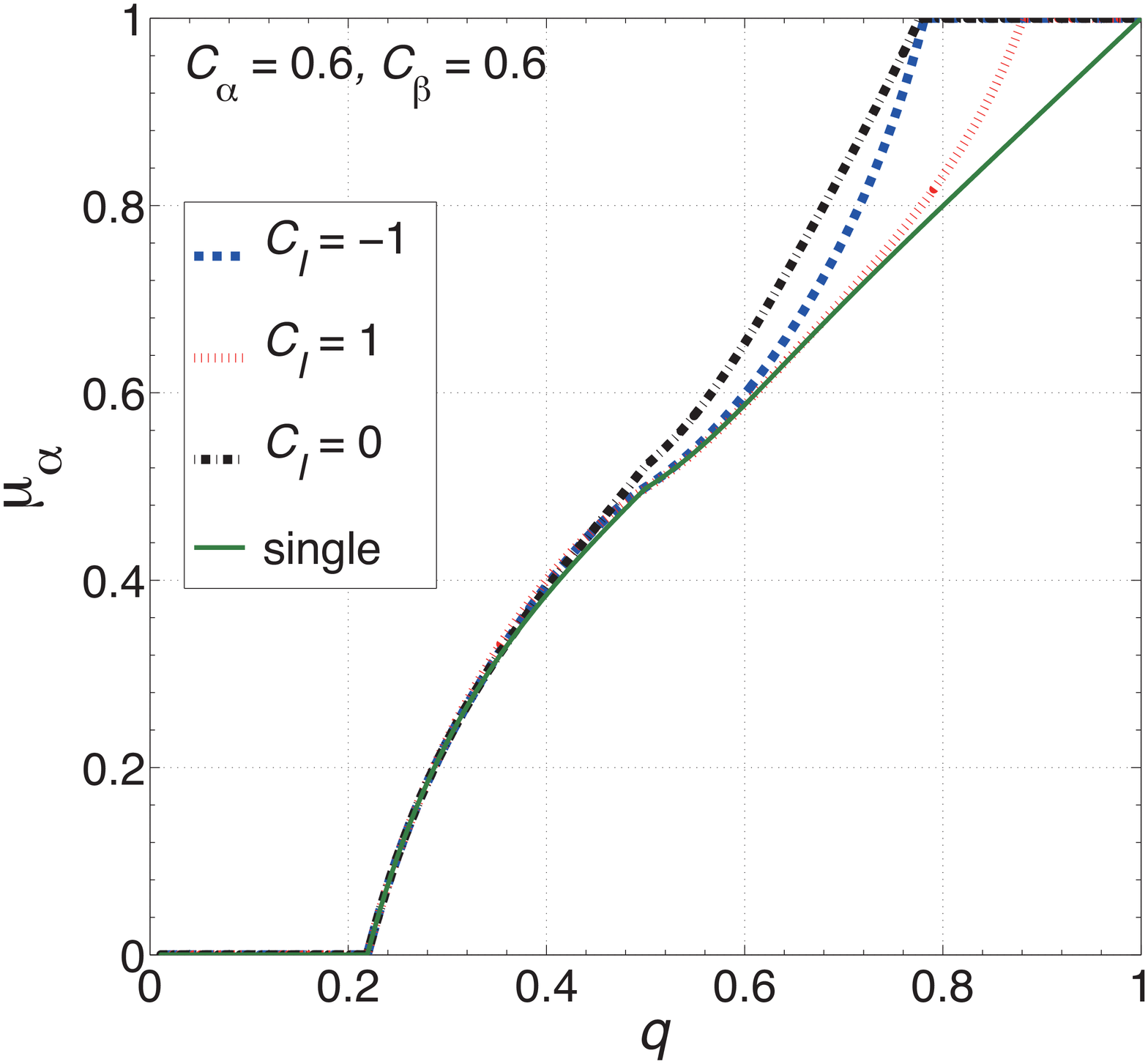}
\includegraphics[width=8cm,height=7cm]{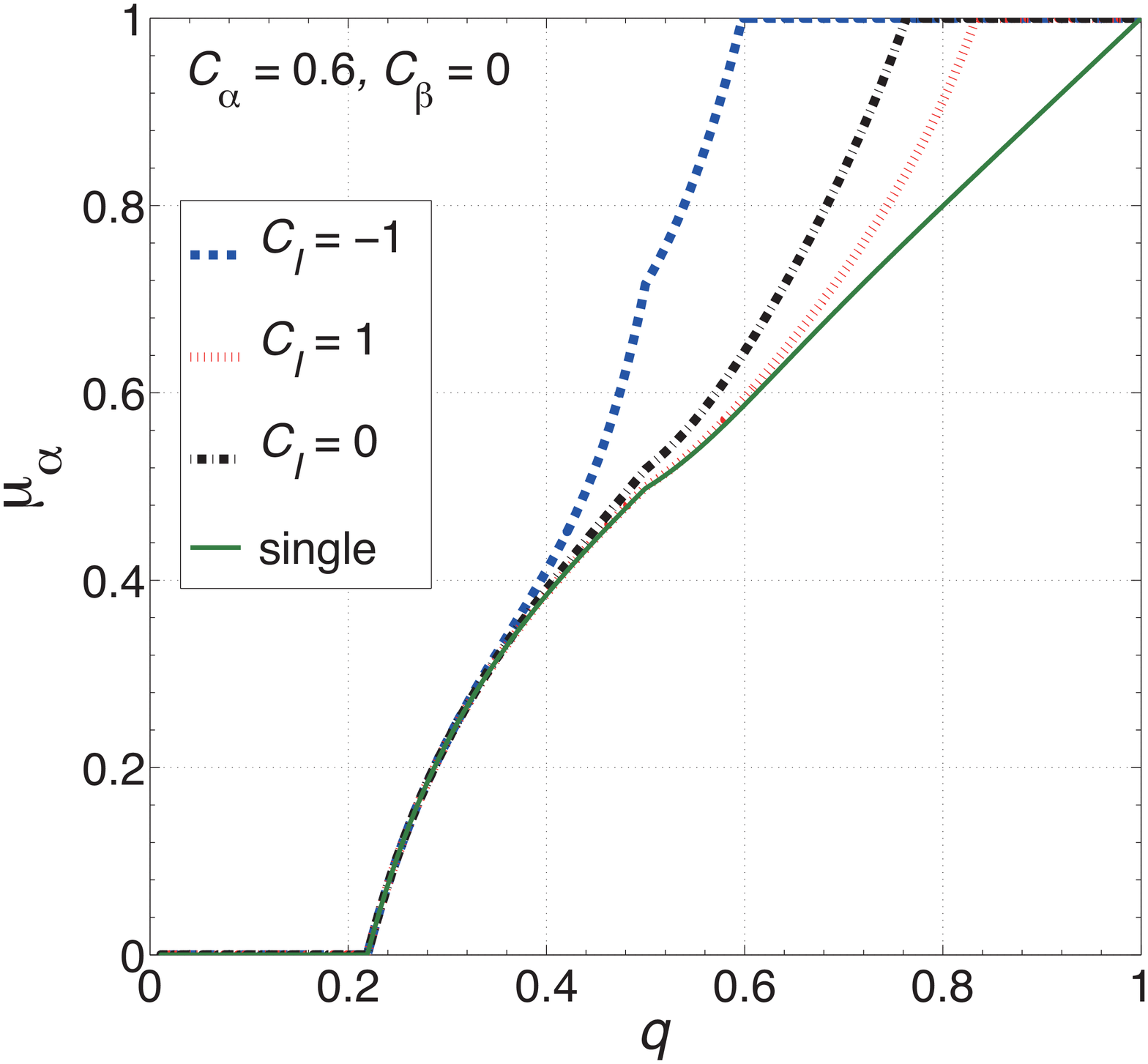}
\includegraphics[width=8cm,height=7cm]{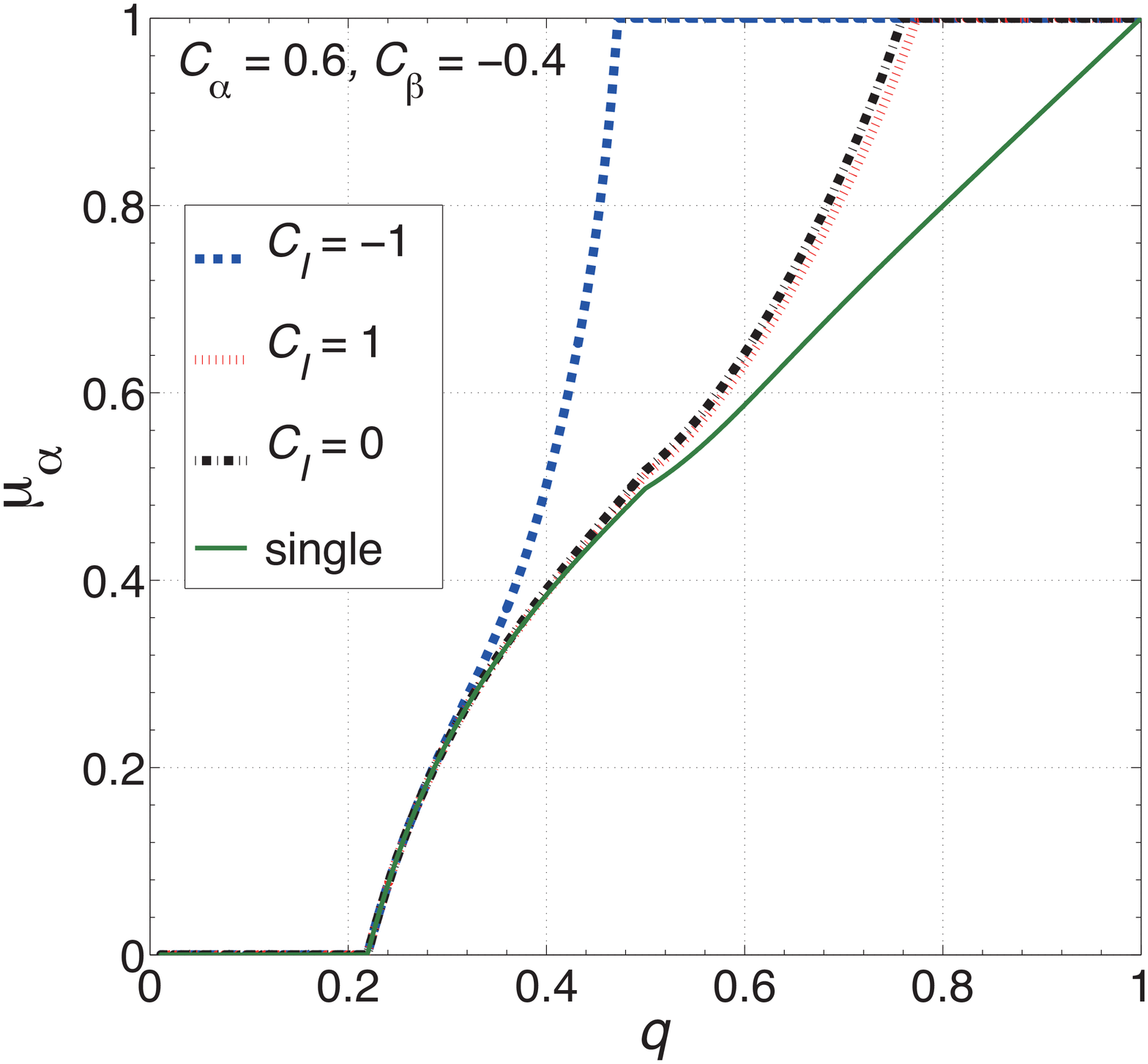}
\end{center}
\caption{(Color on-line)  GC size in layer $\alpha$ versus the initial parameter $q$ for the case where antagonistic networks suffer from TAs, the remaining effect of which is Case F. The GC size in  layer $\alpha$ at stage $t=1$ is also shown, which is the result of the percolation process that is completed in the single layer $\alpha$. The intralayer correlations in layer $\alpha$ are fixed to be positive, {\it e.g.}, $C_\alpha=0.6$, to focus on the effect of intralayer degree-degree correlations in layer $\beta$ and interlayer degree-degree correlations  on the robustness of layer $\alpha$.
\label{Fig: alpha-p}}
\end{figure}

\begin{figure}[htp]
\begin{center}
\includegraphics[width=8cm,height=7cm]{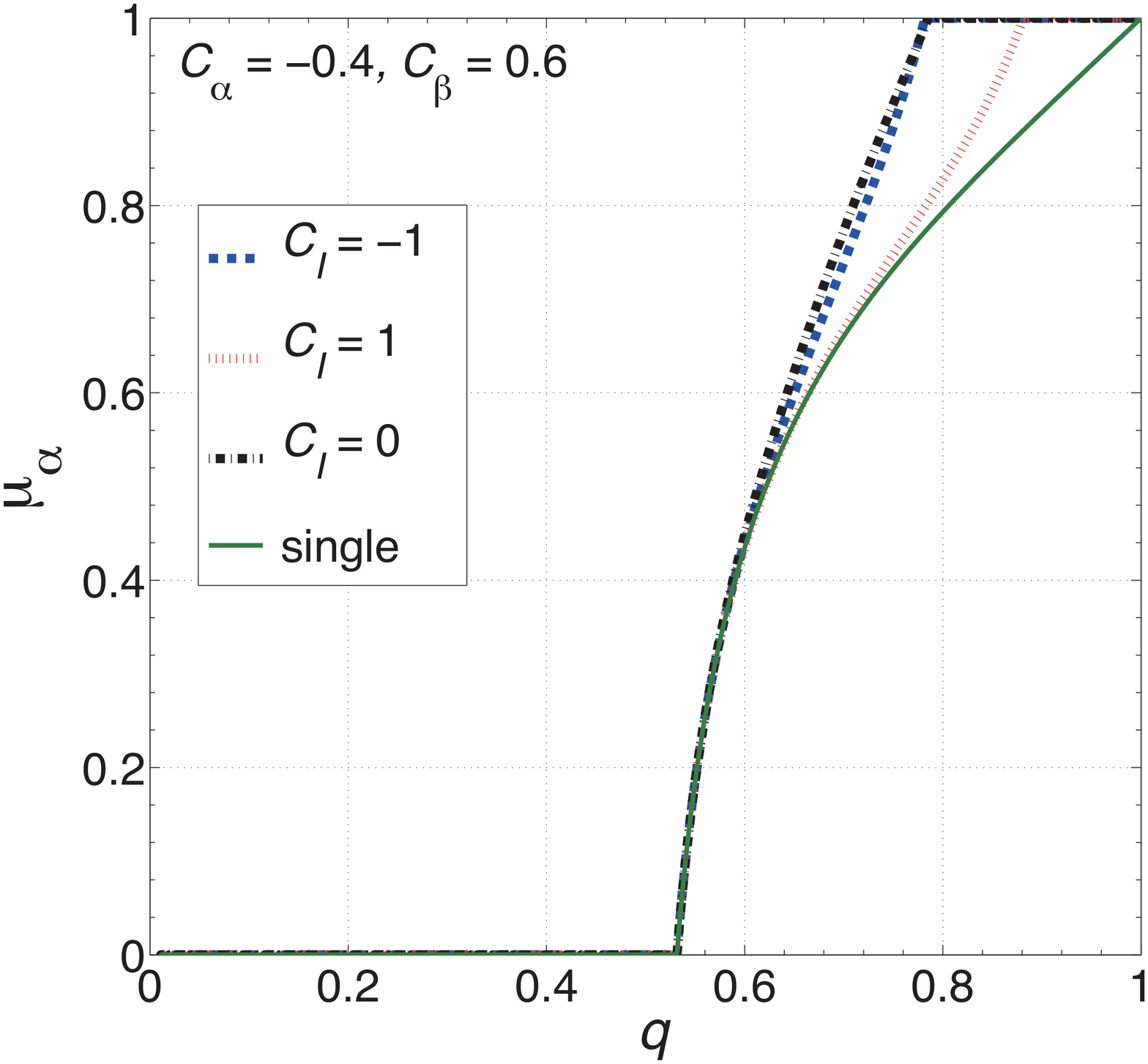}
\includegraphics[width=8cm,height=7cm]{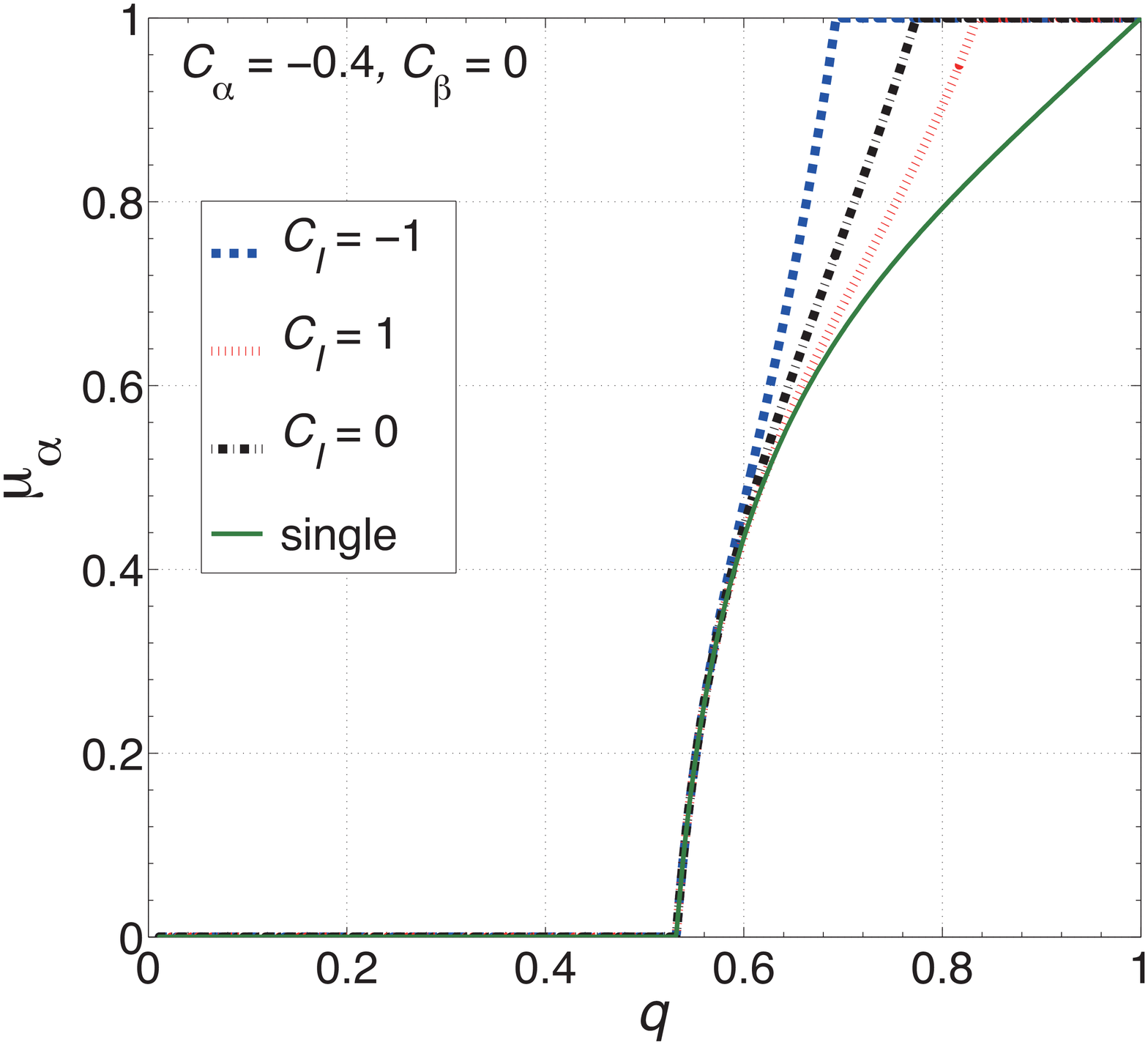}
\includegraphics[width=8cm,height=7cm]{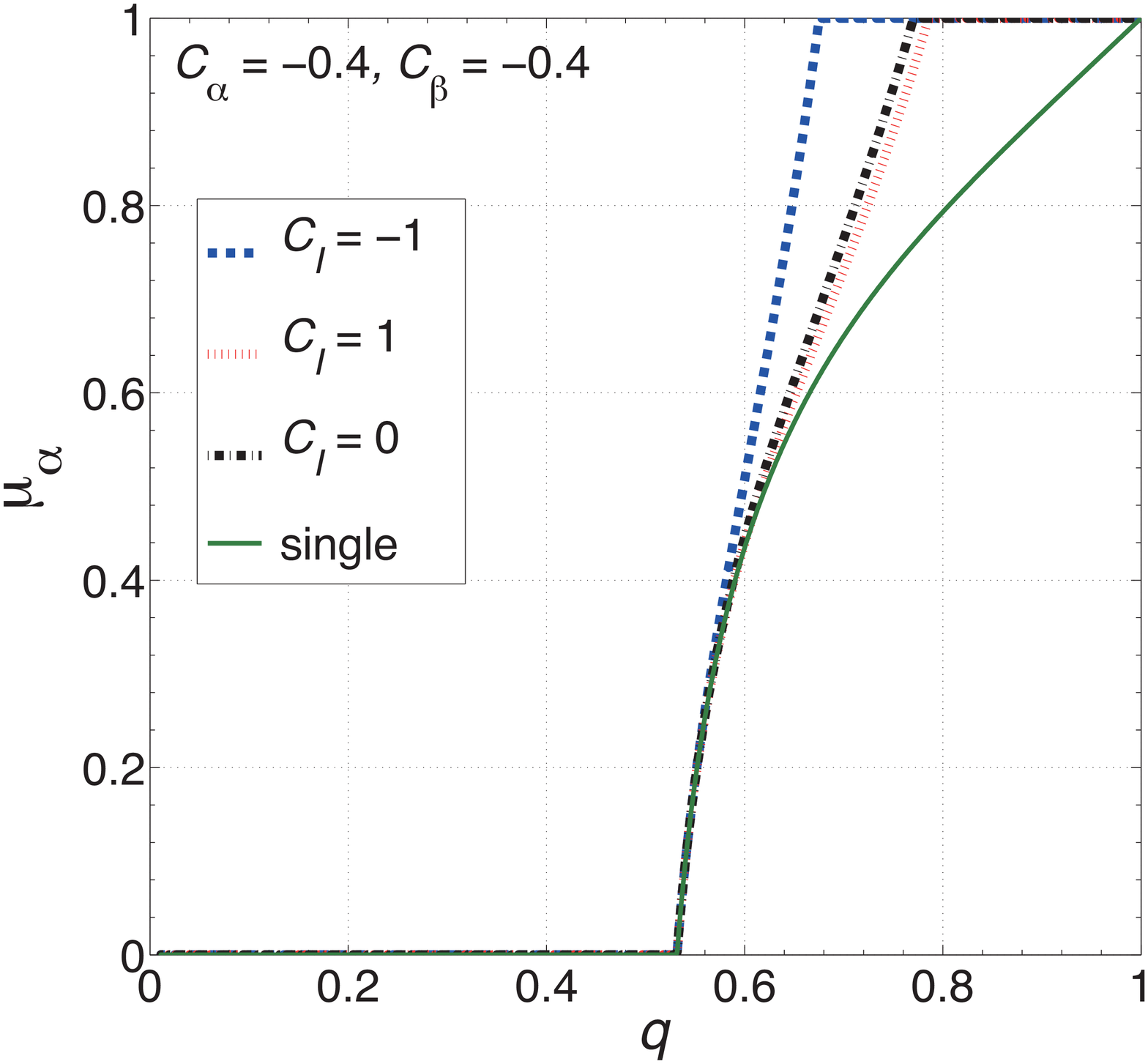}
\end{center}
\caption{(Color on-line) GC size in layer $\alpha$ versus the initial parameter $q$ for the case where antagonistic networks suffer from TAs, the remaining effect of which  is Case F.  The GC size in  layer $\alpha$ at stage $t=1$ is also shown, which is the result of the percolation process that is completed in a single network.  The intralayer correlations in layer $\alpha$ are fixed to be negative, {\it e.g.}, $C_\alpha=-0.4$, to focus on the effect of intralayer degree-degree correlations in layer $\beta$ and interlayer degree-degree correlations  on the robustness of layer $\alpha$.
\label{Fig: alpha-n}}
\end{figure}

We here argue the influence of interlayer and intralayer degree-degree correlations on the robustness of each layer, the specific topologies of which are defined in  \ref{section: procedure}.
We narrow an argument to the Case F and TAs, because 
the percolation processes in each layer in Case Q can be reduced to those of a single network and the influence of degree-degree-correlations on robustness in RFs is considerably smaller than that in TAs.
In Figs. \ref{Fig: alpha-p} and \ref{Fig: alpha-n}, we show examples of the robustness of layer $\alpha$ and the effects of various interlayer and intralayer degree-degree correlations.

We focus on the effect of degree-degree correlations on critical (minimum) robustness ($\mu_{\alpha} \approx 0$) and maximum robustness ($\mu_{\alpha} \approx 1$) of layer $\alpha$, respectively. 
The  thresholds at which $\mu_{\alpha}$ vanishes depend on only its intralayer degree-degree correlations, which are characterized with the Pearson coefficient, $\CA$.
This is because the GC at $t=1$ plays a role of the {\it core} of the final GC and thus if there exists no GC at $t=1$, no node can belong to the GC thereafter.  
On the while, the maximum robustness of layer $\alpha$ is affected from both intralayer degree-degree-correlations  and interlayer degree-degree correlations, the transition point of which accords with  that  for critical robustness of layer $\beta$.  

Let us consider the advantageous conditions for maximum robustness of layer $\alpha$, the parameter example of which  is  $\CA=0.6, \CB = -0.4, C_{I} = -1$ in Fig. \ref{Fig: alpha-p}. 
 Layer $\alpha$ is more robust against TAs if its degree-degree correlations are positive. 
In addition, layer $\alpha$ is more robust if  layer $\beta$ is more fragile due to antagonistic properties. 
 Layer $\beta$ is the most fragile if its intralayer degree-degree correlations are negative and it suffers from TAs, which are realized in the cases where interlayer degree-degree correlations are highly negative.  In this case, nodes of higher degree in layer $\beta$ become inactive because they connect with nodes of lower degree in layer $\alpha$ that are tend to be active due to TAs in layer $\alpha$.  
 
Considering the above, it is  natural that maximum robustness of layer $\alpha$ is the most fragile if the parameter example is  $\CA= -0.4, \CB = 0.6, C_{I} = 1$  in Fig. \ref{Fig: alpha-n}, because each sign of the parameter is opposite with that in the conditions that are advantageous for layer $\alpha$.

\subsection{Possible relevance to real world systems}
As for significance to real world systems, 
the antagonistic networks may serve as a model of complex ecological interactions 
between endangered species and invaders.
Refs. \cite{Watari, Fukasawa} report such ecological relationship in the Amami islands in Japan:
populations of endangered species, such as the Amami rabbit ({\it Pentalagus furnessi}) and  the Amami Ishikawa\rq{}s frog ({\it Odorrana splendida})
were restored to the almost original level at the areas  where  
invaders, such as  the small Indian mongoose ({\it Herpestes auropunctatus}) were exterminated, whereas few were observed in the places where the invaders were established.

Our analysis shows that if initially damaged nodes can be reactivated (Case F), 
the GC size is restored to the original level owing to the antagonistic inhibition to 
the replica nodes as long as the fraction of the initial damage is sufficiently small. 
This is consistent with the above reports. 
The endangered species in the reports have relatively high reproduction rates and short life cycles, 
which may fit the condition of Case F. 
On the other hand, species such as large mammals and primates have low reproduction rates and long life cycles, 
and may correspond to Case Q, for which populations of the species cannot be restored only by the 
antagonistic interaction. 
However, our analysis, in conjunction with Refs. \cite{Watari, Fukasawa},  
implies that, even in such cases, combination of 
increasing the reactivation rate of endangered species 
by human-induced methods such as relocation and extermination of invasive species
is an effective scheme for conserving ecological systems. 

\section{Summary}
In this paper, we  developed an analytical methodology based on the cavity method to  study the  robustness of duplex networks  coupled with antagonistic interlinks, considering intralayer and interlayer degree-degree correlations. 
We investigated two scenarios according to whether initially failed  nodes are able to revive (Case F) or not (Case Q) with the aid of their replica nodes.
 In both Cases Q and F, we showed that the failure process periodically repeated because of the peculiarity of the antagonistic property of interlinks and the percolation transition  exhibited continuous.
 The oscillation was due to hysteresis of each layer, which led to the inconsistency between theory and experiment 
 particularly for Case F. Therefore, we introduced a heuristic treatment 
 for improving the theoretical prediction accuracy, the utility of which was verified by numerical experiments.  

We also argued the most advantageous situations for layer $\alpha$ in terms of degree-degree correlations, employing bimodal networks.
While the minimum robustness of layer $\alpha$ ($\mu_{\alpha} \approx 0$) was affected from only intralayer degree-degree correlations in layer $\alpha$, the  maximum robustness of layer $\alpha$ ($\mu_{\alpha} \approx 1$) was influenced from various degree-degree correlations;
 the most robust  situations are positive intralayer degree-degree correlations  in layer $\alpha$ and negative intralayer degree-degree correlations  in layer $\beta$ and  negative interlayer degree-degree correlations. 
As for significance to real world systems, a possible relevance to ecological systems that are composed of 
endangered and invasive species is mentioned. 

 
Future works include to  construct  model and analytical framework that works with more realistic  settings by network approach.

\section*{Acknowledgment}
This work was partially supported by KAKENHI No. 25120013 (YK). 
\newpage
\appendix
\section{Reconfirming the periodicity by the heuristic}
\label{section: reconfirming}
Although we  already made sure that the percolation processes oscillates in both Cases Q and F in Sec. \ref{section: process},  
we here reconfirm these  by  the heuristic in Sec. \ref{section: heuristic(micro)}.

\subsection{Case Q}   
 The active variable of each node at stage $t=3$ is provided as
\begin{eqnarray}
s^{t=3}_{\iA}  {\rm (Q)}&=&\sone m^{t=3}_{p\to \iA}\cr
&=&
(\sone)^2+\sone\left(1-\sone\right)\prod_{\aB \in \partial \iB} m^{t=2}_{\aB \to \iB}\cr
&-&  {(\sone)}^2 \prod_{\aA \in \partial \iA} m^{t=1}_{\aA \to \iA}
\left(1-\prod_{\aB \in \partial \iB} m^{t=2}_{\aB \to \iB}\right). \cr
&=&\sone\left(1 - \hspace{-2mm}\prod_{\aA \in \partial \iA} \hspace{-1mm}m^{t=1}_{\aA \to \iA}
\left(1-\hspace{-2mm}\prod_{\aB \in \partial \iB} m^{t=2}_{\aB \to \iB}\right)\hspace{-1mm}\right).\cr
&&\label{siA3Q}
\end{eqnarray}
 Substituting $m^{t=1}_{\aA \to \iA}=0$ in  Eq. (\ref{siA3Q}),
we obtain $s^{\ast t=3}_{\iA}  {\rm (Q)} = \sone$,
which also shows that $m^{\ast t=3}_{\aA \to \iA}$ is equivalent with $m^{t=1}_{\aA \to \iA}$.
Using  $s^{\ast t=3}_{\iA}  {\rm (Q)}$ and $m^{ \ast t=3}_{\aA \to \iA}$, we obtain the result  on the GC size,  namely
\begin{eqnarray}
\sigma^{t=3}_{\iA}
&\approx&s^{t=3}_{\iA} {\rm (Q)} \left(1-\Mcthree\right)\cr
&=&\sone\left(1 - \hspace{-2mm}\prod_{\aA \in \partial \iA} \hspace{-1mm}m^{t=1}_{\aA \to \iA}\right)^2
+ \cr && \prod_{\aA \in \partial \iA} \hspace{-1mm}m^{t=1}_{\aA \to \iA} \left(1-\hspace{-2mm}\prod_{\aA \in \partial \iA} \hspace{-1mm}m^{t=1}_{\aA \to \iA}\hspace{-1mm}\right) \Mtwo\cr
&=&\sone \left(1-\Mone\right)\cr
&=& \sigma^{t=1}_{\iA}.
\label{g3Qmod}
\end{eqnarray}

 \subsection{Case F}   
Substituting Eq. (\ref{s3F1}) to Eq. (\ref{s_iB4 (F)}),
we describe  $s^{t=4}_{\iB} {\rm (F)}$  in detail,
\begin{eqnarray}
s^{t=4}_{\iB} {\rm (F)} 
&=&1-s^{\ast t=3}_{\iA}\left(1-\Mthree\right)\cr
&+&\sone\left(\Mone-\Mthree\right)\cr
&\cdot&\left(1-\Mtwo\right).
\label{s_iB4 (F)2}
\end{eqnarray}
Note that  $\Mone\Mthree$ is replaced with $\Mthree$  because of Table \ref{table1}.
Supposing that $\Mtwo$ vanishes in Eq. (\ref{s_iB4 (F)2}), which also replaces $s^{\ast t=3}_{\iA}$ with $s^{t=1}_{\iA}$ because of Eq. (\ref{s3mod}),  we define 
\begin{eqnarray}
s^{\ast t=4}_{\iB} {\rm (F)} 
&\equiv&1-s^{t=1}_{\iA}\left(1-\Mthree\right)\cr
&+&\sone\left(\Mone -\Mthree\right)\cr
&=&1-s^{t=1}_{\iA}+\sone\Mone\cr
&=&s^{t=2}_{\iB}  
\label{s_iB4 (F)3}, 
\end{eqnarray}
where the last equal sign is because of Eq. (\ref{siB}).
Employing $s^{\ast t=4}_{\iB} {\rm (F)}$ instead of  $s^{t=2}_{\iB}$  in Eq. (\ref{hstep_B}) and (\ref{vstep_B}), we derive the
provisional message at $t=4$, $m^{\ast t=4}_{\aB \to \iB}$, which completely accords with  $m^{t=2}_{\aB \to \iB}$ because of Eq. (\ref{s_iB4 (F)3}).

 The label of the GC at $t=4$ is derived using the original interlayer message $s^{t=4}_{\iB} {\rm (F)}$ and the provisional message $m^{\ast t=4}_{\aB \to \iB}$. Namely,
\begin{eqnarray}
\sigma^{t=4}_{\iB}
&\approx&s^{t=4}_{\iB} {\rm (F)} \left(1-\Mcfour\right)\cr
&=&s^{t=4}_{\iB} {\rm (F)} \left(1-\Mtwo\right)\cr
&=&s^{t=2}_{\iA} \left(1-\Mtwo\right)\cr
&-& \left(1-\sone\right)\left(1-\Mthree\right)\cr
&\cdot&\Mtwo\left(1-\Mtwo\right)\cr
&=& \sigma^{t=2}_{\iB},
\label{g4mod}
\end{eqnarray}
where the last equal sign is derived because $\Mtwo\left(1-\Mtwo\right)$ always vanishes.
We also note that the label of the provisional GC,  which is defined as
$\sigma^{\ast t=4}_{\iB} \equiv s^{\ast t=4}_{\iB} {\rm (F)}\left(1-\Mcfour\right) = \sigma^{t=2}_{\iB}$, also correctly evaluates the GC at $t=4$,  because there exist no nodes that incorrectly belong to the GC at $t=4$ even if we set   $s^{\ast t=4}_{\iB} {\rm (F)}$ instead of  $s^{t=4}_{\iB} {\rm (F)}$.

\section{Notations}
\label{section: notations}
\begin{tabularx}{80mm}{|l|X|}
\hline
Indice& Definitions or meanings\\ \hline
$\alpha$, $\beta$ & Layer $\alpha$ and layer $\beta$, respectively. \\
$\aA$& Index of a  function node on a link in layer $\alpha$.\\
$i_\alpha$, $j_\alpha$ & Index of a node (variable node) in layer $\alpha$.\\
$p$ &  Index of a function node on an  interlink.\\
$\partial \iA$ & Set of nodes  in layer $\alpha$ that connect with  node $\iA$.\\
$|\partial \iA|$ & Degree of   node $\iA$.\\
$\kA$, $\lA$ & Degree of a node in layer $\alpha$.\\
$P(\kA,\kB)$ & Interlayer joint degree distribution.\\
$\rA(\kA)$ & Distribution of the degree of a node in one terminal, given a randomly chosen intralink in layer $\alpha$.\\ 
$\rA(\kA,\kB|\lA, \lB)$ &  Conditional interlayer degree-degree distribution.
\\ 
$\CA, \CB,\CI$ &  Pearson coefficients in layer $\alpha$,  layer $\beta$,  and  between layers, respectively.\\
Case Q, Case F & Remaining effect of the initial failure is quenched  and free, respectively.\\
\hline
$\psi_{\iA}$ &  Binary variable that represents whether a node $\iA$ initially fails ($\psi_{\iA}=0$) or not ($\psi_{\iA}=1$).  \\
$s^{2t\rq{}-1}_{\iA}$ &  Binary variable that represents whether a node $\iA$  fails ($s^{2t\rq{}-1}_{\iA}=0$) or not  ($s^{2t\rq{}-1}_{\iA}=1$)  at the onset of the stage $2t\rq{}-1$.
\\
$m^{2t\rq{}-1}_{\aA\to\iA}$  & Message that propagates from function node $\aA$ to (variable) node $\iA$ at stage ${2t\rq{}-1}$, where $\partial \aA=\{\iA ,\jA \}$. If node $\jA$ belongs to the GC on $\iA$-cavity system, $m^{2t\rq{}-1}_{a\to\iA}=0$ is completed; otherwise $m^{2t\rq{}-1}_{a\to\iA}=1$ is completed.  
\\
$\sigma^{2t\rq{}-1}_{\iA} $ &  Binary variable that represents whether  node $\iA$  belongs to the GC in layer $\alpha$ ($\sigma^{2t\rq{}-1}_{\iA}=1$) or not ($\sigma^{2t\rq{}-1}_{\iA}=0$) at stage $2t\rq{}-1$.
\\
\hline
$q$ & Fraction of (variable) nodes that are active (in other words, have not failed) at the onset of the initial stage. \\
$\qA$& Active probability of (variable) nodes in layer $\alpha$  at the onset of the stage $2t\rq{}-1$, the degrees of which are $\kA$; the degrees of its replica nodes are  $\kB$. \\ 
$I^{\alpha, {t=1}}_{\lA,\lB}$  &  Message on an intralink in layer $\alpha$ at  stage $t=1$ from the macroscopic viewpoint,  which is defined in Eq. (\ref{IAdef}). 
\\
$\mu^{2t\rq{}-1}_{\alpha} $ &  Expected  GC size in layer $\alpha$  at the end of stage $2t\rq{}-1$.
\\
\hline
\end{tabularx}

\begin{thebibliography}{99}
\bibitem{we}
The  fraction of nodes that are incorrectly evaluated as the constituents of the GC at $t=3$ in Case F, which are marked with star  in Fig. \ref{transitionmod}, is  evaluated 0.02 at most  in our study  for duplex networks, each layer of which is regular random networks in which degree of node is $4$.
\end{thebibliography}
\end{document}